\documentclass[12pt, draftclsnofoot, onecolumn]{IEEEtran}
 \usepackage[english]{babel}
 \usepackage{fancyhdr}
\usepackage{epsfig}
\usepackage{threeparttable}
\usepackage{epsf,epsfig}
\usepackage{amsthm}
\usepackage{amsmath}
\usepackage{amssymb}
\usepackage{amsfonts}
\usepackage[noadjust]{cite}
\usepackage{dsfont}
\usepackage{graphicx,amssymb,amsmath}
\usepackage{color}
\usepackage{soul}
\usepackage{algorithmic,algorithm}
\usepackage{url}
\usepackage{subcaption}

\newtheorem{lemma}{Lemma}

\newtheorem{proposition}{Proposition}

\newtheorem{remark}{\bf Remark}

\def\proof{\noindent{\emph{Proof:} }}

\def\phi{\varphi}

\def\l{\left}
\def\r{\right}
\def\({\left(}
\def\){\right)}

\def\b0{{\mathbf{0}}}

\newcommand{\tr}{\mathrm{tr}}

\begin{document}

\title{\huge Reduced-Dimension Design of MIMO Over-the-Air Computing for Data Aggregation in  Clustered IoT Networks}
\author{Dingzhu Wen, Guangxu Zhu, and Kaibin Huang     \thanks{\setlength{\baselineskip}{13pt} \noindent D. Wen, G. Zhu, and K. Huang are with the Dept. of Electrical and Electronic Engineering at The  University of  Hong Kong, Hong Kong (Email: huangkb@eee.hku.hk). }}
\maketitle

\vspace{-18mm}
\begin{abstract}
One basic operation of  \emph{Internet-of-Things} (IoT) networks is aggregating distributed sensing data collected over wireless channels to compute a desired function, called \emph{wireless data aggregation} (WDA). In the presence of dense sensors, low-latency WDA poses a design challenge for high-mobility or mission critical IoT applications. A technology called \emph{over-the-air computing} (AirComp) can dramatically reduce the WDA  latency  by  aggregating distributed data ``over-the-air" using the waveform-superposition property of a multi-access channel. In this work, we design \emph{multiple-input-multiple-output} (MIMO) AirComp for computing a \emph{vector-valued function} in a clustered  IoT network with  multi-antenna sensors forming clusters and a multi-antenna \emph{access point} (AP) performing WDA. The resultant high-dimensional but low-rank MIMO channels makes it important to reduce channel/signal dimensionality in AirComp to avoid exposure to noise from channel null-spaces. The design challenge lies in the integration of simultaneous dimension-reduction and joint-equalization (without decoupling) of many MIMO channels with correlation and heterogeneous ranks. By tackling  the challenge, we develop in this work  a framework of reduced-dimension MIMO AirComp. The key component is \emph{decomposed aggregation beamforming} (DAB) for the AP. Consider the case of \emph{separable} channel clusters with non-overlapping \emph{angle-of-arrival} (AoA) ranges. The optimal DAB is proved to have the architecture where \emph{inner components} extract the dominant eigen-spaces of corresponding channel clusters and \emph{outer components} jointly equalize the resultant low-dimensional channels. Consider the more complex case of \emph{inseparable}  clusters. We propose a suboptimal DAB design where the \emph{inner component} performs both dimension reduction and joint equalization over clustered-channel covariance matrices and the \emph{outer component} jointly equalizes the small-scale fading channels. As part of the said framework, we also design efficient algorithms for rank optimization of individual DAB components  and channel feedback leveraging  the AirComp principle. The proposed framework is shown by simulation to substantially reduce AirComp error compared with the existing design without considering channel structures.

\end{abstract}

\section{Introduction}
\begin{figure}[t]
    \centering
    \begin{subfigure}[b]{0.47\textwidth}
        \includegraphics[width=\textwidth]{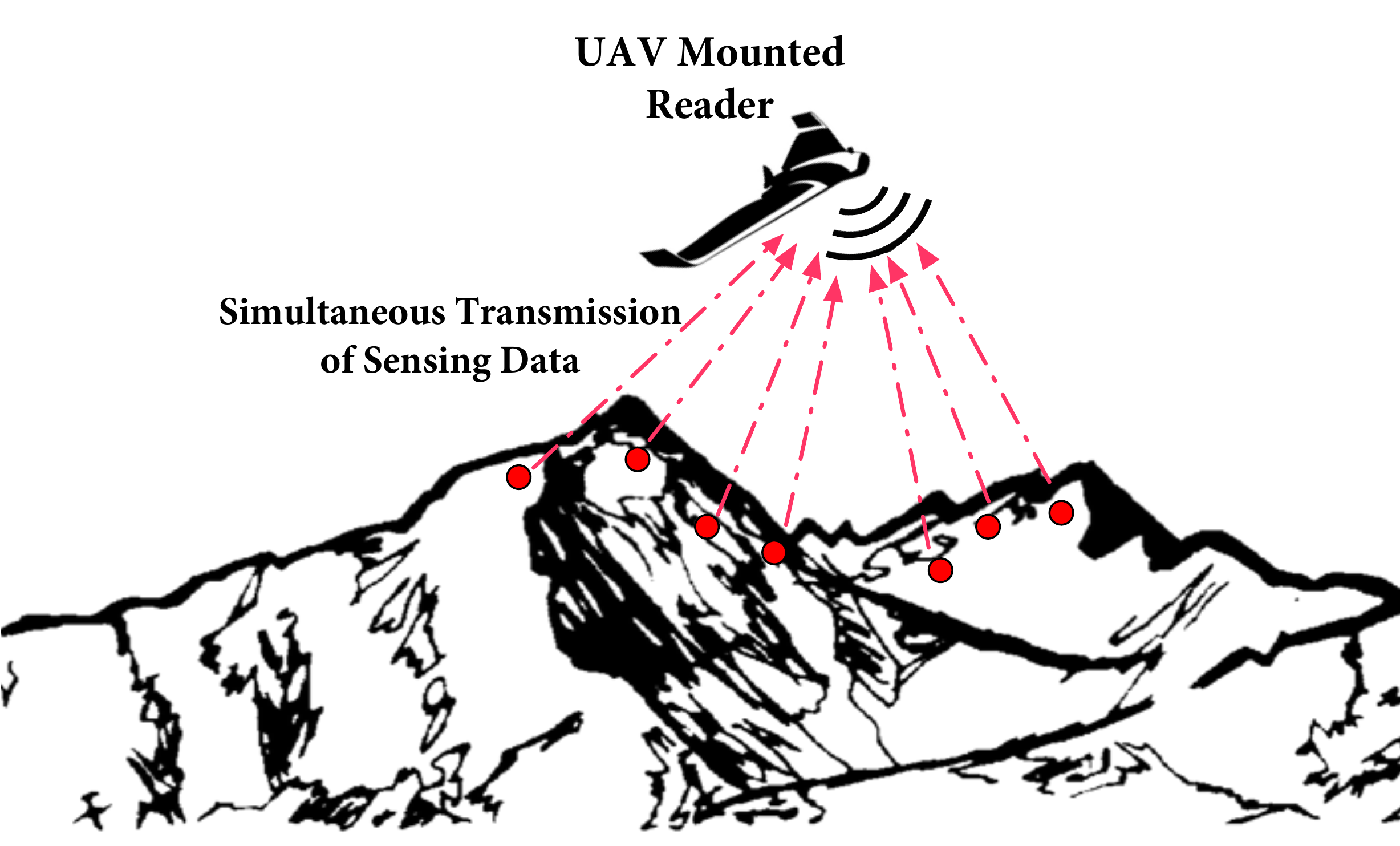}
        \caption{Wireless data aggregation in wild sensor networks with high mobility AP.}
    \end{subfigure}
    ~ 
    \begin{subfigure}[b]{0.47\textwidth}
        \includegraphics[width=\textwidth]{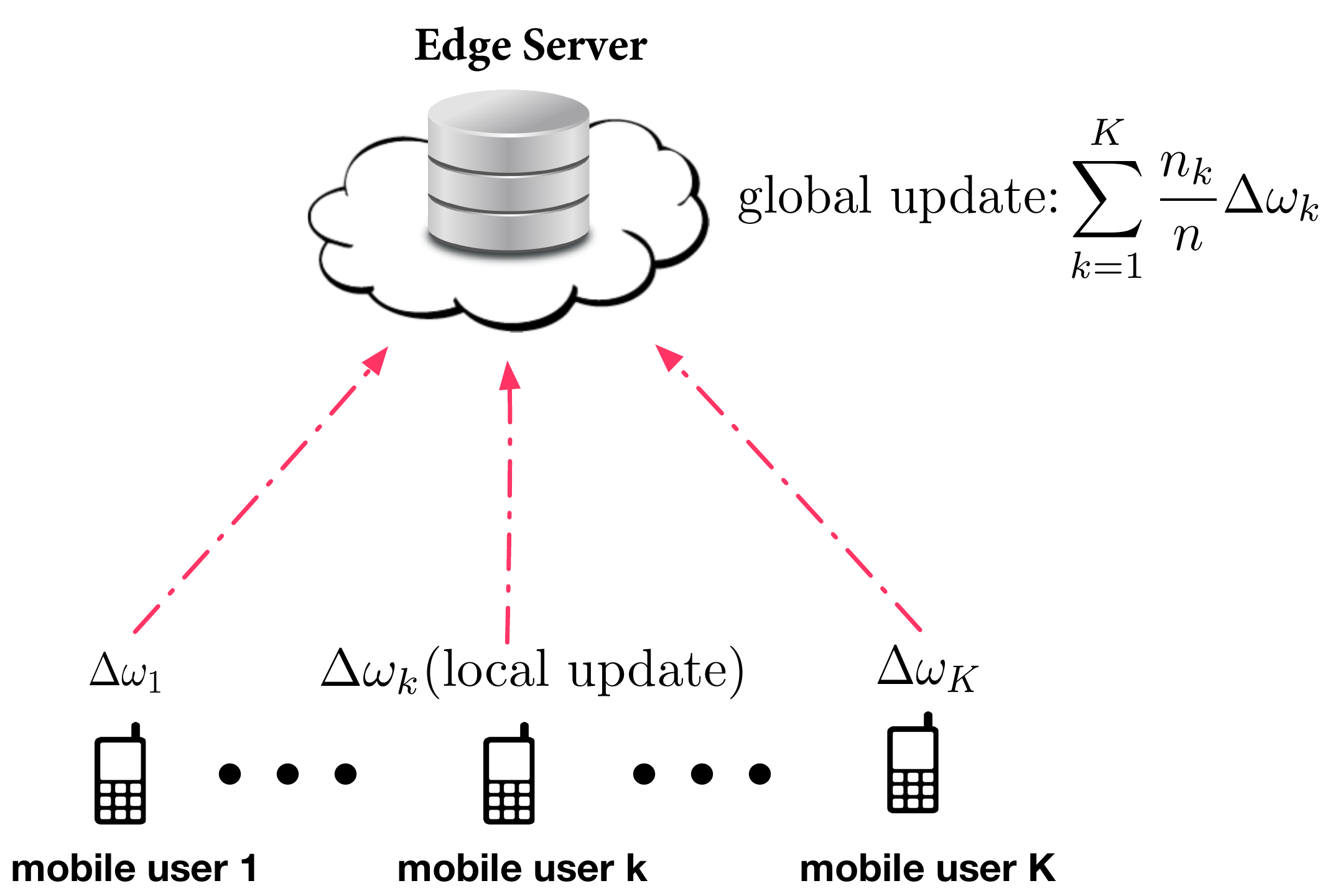}
        \caption{Wireless data aggregation for federated learning.}
    \end{subfigure}
    \caption{Two examples of fast WDA in IoT networks.}\label{fig:WDAapp}
\end{figure}

The future  \emph{Internet-of-Things} (IoT) will collect distributed data from an enormous number of edge devices (sensors and smartphones), perform computation and inference using the data, and then use equally many actuators to control the physical environment \cite{iot}. Thereby, IoT is expected  to automate various operations of our society such as manufacturing, heathcare, and traffic control. Among others, one main challenge of designing  IoT networks is fast \emph{wireless data aggregation} (WDA), referring to  fast collection of  distributed data from edge devices via wireless transmission. The challenge arises in scenarios characterized by many devices, high mobility or heavy data uploading. One example of high-mobility WDA is data collection using a UAV-mounted reader \cite{UAV} and another example of heavy data uploading is federated machine learning \cite{federatedLearning}, both of which are illustrated in Fig. \ref{fig:WDAapp}. The ultra-low latency requirement of fast WDA cannot be met by the traditional ``transmit-then-compute" approach of designing an air interface that incurs unacceptable transmission latency in the said scenarios. A more efficient design approach is ``transmit-and-compute" that integrates transmission and computation. A specific vein of research based on  this approach is called  \emph{over-the-air computation} (AirComp), which  attracts increasing research interests recently \cite{Guangxu,Xiaoyang,AirComp2018}. The principle of AirComp is to exploits co-channel interference  for computing a function of distributed data, thereby allowing simultaneous transmission and dramatic latency reduction. The particular class of functions exactly computable using AirComp is called
\emph{nomographic functions} \cite{aircomp1,aircomp2}, that have the following form:
\begin{equation}\label{eq:AirComp}
\bar{\bf Z} = q\bigg(\sum\limits_kf_k({\bf Z_k})\bigg),
\end{equation}
where ${\bf Z_k}$, $f_k$, and $q$ are the input data of the $k$-th device, the corresponding pre-processing function, and the post-processing function, respectively. In \cite{NomF1,NomF2}, it is proved that an arbitrary function can be decomposed as the sum of nomographic functions, which is thus Air-Computable.

In next-generation massive  \emph{multiple-input and multiple-output} (MIMO) IoT networks, large-scale antenna arrays will support AirComp of vector-valued functions, called \emph{MIMO AirComp}. Furthermore, the high-resolution arrays are capable of resolving  sensors into clusters \cite{JSDM,Gesbert,Lau}. Exploiting the structure of the resultant clustered channels can reduce noise and facilitate joint channel equalization (without decoupling) in AirComp, thereby reducing its errors as well as channel-feedback overhead. This motivates the current work on developing the framework of reduced-dimension MIMO AirComp.

\subsection{Wireless Data Aggregation  by AirComp}
Consider WDA by AirComp in a multi-access channel where an  \emph{access point} (AP) aims at obtaining the desired functional value in \eqref{eq:AirComp} with minimum distortion. The original idea of AirComp appeared in \cite{aircomp2}. The design relies on structured codes (i.e., lattice codes) to cope with channel distortion introduced by the multi-access channel. It was subsequently discovered in \cite{aircomp3} that simple analog transmission without coding but with channel pre-equalization can achieve the minimum distortion if the  data sources are \emph{independent and identically distributed} (i.i.d.) Gaussian. If this assumption does not hold, coding can be still beneficial e.g., as shown in \cite{aircomp7} for the scenario where data sources follow the  bivariate Gaussian distribution \cite{aircomp7}. Nevertheless, the simplicity of the optimal design for the Gaussian case  has inspired a series of follow-up research on making AirComp practical \cite{aircomp4, aircomp5, aircomp6}.  By measuring the AirComp distortion using  \emph{mean squared error} (MSE), the optimal power allocation and  outage performance under a distortion constraint  are studied in \cite{aircomp4} and \cite{aircomp5}, respectively. The implementation of AirComp typically requires CSI at transmitters for channel pre-equalization. An attempt to relax the requirement was made in \cite{aircomp6} where randomized transmission without CSI realizes AirComp at the cost of increased latency. Another practical issue for implementing  AirComp is synchronizing the transmission of edge devices. One design addressing this issue is proposed  in \cite{aircomp8} that modulates the data into transmit power to relax the synchronization requirement. As a result, only coarse block-synchronization is required for realizing AirComp. An alternative scheme, called AirShare, is to broadcast  a shared clock to all devices  \cite{aircomp9}.

The prior work described above focuses on AirComp of scalar-valued functions. Most recent research in the area aims at MIMO AirComp using MIMO techniques to enable vector-valued functional computation  \cite{Guangxu}. In particular, receive beamforming targeting WDA, called \emph{aggregation beamforming}, is proposed in \cite{Guangxu} to compute vector-valued functions   by spatial multiplexing and reduce AirComp distortion by spatial diversity. Along the same vein, the current work targets clustered IoT networks and addresses the issue of how to exploit channel structure for improving the performance of MIMO AirComp.

Last, while AirComp is mostly deployed in  computation-centric networks as discussed above, it is worth mentioning that the AirComp operation has been also leveraged in  rate-maximization schemes   such as two-way relay  \cite{aircompRe} and MIMO lattice decoding \cite{aircompLat}.

\subsection{Reduced-Dimension Design for  Massive MIMO Systems}

In next-generation wireless systems, large-scale antenna arrays are expected to be deployed at APs (each with hundreds to thousands of elements) and mobile devices (each with tens of elements) \cite{LarssonMIMO}. In such massive MIMO systems, one research focus is to reduce complexity in transceiver designs and  thereby also reduce  overhead for  CSI feedback. There exist a rich literature of such designs  \cite{MIMOSurvey}. The ``phased-zero-forcing (ZF)" precoding scheme proposed in \cite{PZF} achieves complexity reduction by combining ZF precoding in the baseband domain and phase control in the \emph{radio frequency} (RF) domain.  On the other hand, a hierarchical architecture for implementing multiuser ZF receiver based on user clustering is shown in \cite{NestedZF1,NestedZF2} to yield complexity reduction. Another popular approach for reduced-dimension MIMO is called \emph{hybrid beamforming} that decompose a MIMO transceiver into two cascaded components for analog and digital implementation \cite{GX_HybB,JZ_HybB,RH_HybB}. For clustered MIMO channels, this implementation based architecture can dramatically reduce the number of required RF chains and the complexity of digital processing \cite{JSDM_MM}.

There exists one more key  approach for  reduced-dimension precoding design for massive MIMO downlink, which is closely related to the current work. The high spatial resolution of a large-scale arrays at an AP makes it possible to resolve the cluster structure embedded in multiuser MIMO channels. The main  principle of the design  approach  is to decompose  each  MIMO channel into a slow-time-scale component, namely its \emph{(spatial) covariance matrix}, and a fast-time-scale component, namely \emph{small-scale fading} \cite{Gesbert, JSDM, Lau, JSDM1, JSDM_NB}. The covariance matrix is jointly determined by array and channel-topology parameters including the  size and  antenna-spacing of the transmit array, and  \emph{angles of arrival} (AoA) and \emph{angular spreads} (AS) of user clusters. The channel decomposition leads to an  efficient hierarchical beamformer structure cascading a slow-time-scale and a fast-time-scale components, which are computed based on the covariance and fading matrices, respectively \cite{JSDM}. The former is high-dimensional but requires infrequent or one-time computation. On the other hand, the latter is low-dimensional and hence supports efficient periodic computation  and CSI feedback. The beamforming structure is proved in  \cite{JSDM} and simultaneously in  \cite{Gesbert} to be capacity-achieving as the transmit-array size grows.  The inspiring result has motivated a series of follow-up research that extends the mentioned beamforming design to  millimeter-wave frequency bands \cite{JSDM_MM},  includes opportunistic user selection \cite{JSDM1},  and considers the \emph{minimum mean-square-error} (MMSE) criterion \cite{JSDM_NB}.

The current work builds on the above prior work to design  \emph{reduced-dimension aggregation beamforming} for MIMO AirComp. In particular, we consider the same model of  clustered massive MIMO channel and the same decomposed beamforming structure as in \cite{Gesbert,JSDM,Lau}. However, prior work targets rate-centric downlink systems and thus the  objective for multiuser beamforming  is \emph{sum-rate maximization}. In contrast, we consider a  computation-centric IoT system and the design criterion for aggregation beamforming is  \emph{minimizing distortion in functional computation}. As a result of different design criteria, the two types of beamforming can be differentiated in two aspects described as follows.
\begin{enumerate}
\item {\bf (DoF usage)} For  multiuser beamforming, the spatial \emph{degrees-of-freedom} (DoF) at the AP  are first allocated for decoupling users' data streams by suppressing inter-user interference; the remaining DoF are then applied to enhancing the reliability of individual streams. As a result, the required number of DoF scales linearly with the number of simultaneous users. In contrast, aggregation beamforming utilizes  all  DoF for  suppressing computation errors via joint multiuser-channel equalization without decoupling them. In other words, user separation is unnecessary and   the aggregation process leverages ``interference" instead of suppressing it \cite{aircompRe}. As a result, AirComp does not incur the said scaling and thus requires far fewer DoF  than the rate-maximization counterpart when the number of users is large.

\item {\bf (User separability)} Multiuser beamforming is infeasible when users  lack spatial separability e.g., in the case of overlapping AoA ranges \cite{Gesbert, JSDM}. In contrast, aggregation beamforming does not require user/channel separability.

\end{enumerate}

The above fundamental differences   pose new challenges in aggregation-beamforming design.

\subsection{Contributions and Organization}

In this paper, we consider WDA  in a clustered massive MIMO  network,  where an AP equipped with a large-scale array performs MIMO  AirComp over distributed transmissions by mobile devices. The existing design of  aggregation beamforming  assuming structureless channels with rich scattering \cite{Guangxu}. Its direct application in the current case  would unnecessarily expose AirComp to strong noise from null spaces of low-dimensional cluster channels. This motivates reduced-dimension aggregation beamforming, whose design faces the following challenges.

\begin{itemize}
\item A naive approach for {\bf reduced-dimension aggregation beamforming} is to use the large-scale receive array to extract and separate low-dimensional signals from the dominant eigen-spaces of different cluster channels, which are then aggregated. First of all, this approach is infeasible when the clusters are inseparable due to overlapping AoA ranges \cite{JSDM}. Even if they are separable, the signals with heterogeneous dimensionality cannot be directly aggregated, and the said approach may not be optimal.

\item Signal-dimension reduction shortens the distances between the resultant channel sub-spaces of different devices, thereby reducing the AirComp error. On the other hand, the operation also reduces received signal power and hence increases the error. Balancing these two effects of signal-dimension reduction gives rise to a new  problem called {\bf channel-rank selection}.

\item {\bf Channel feedback} should exploit channel low-dimensionality and AirComp operation for feedback-overhead reduction.

\end{itemize}

In this work, we attempt to tackle the above challenges. The main contributions of the work are summarized below.

\begin{itemize}
\item {\bf Decomposed Aggregation Beamforming (DAB) for Disjoint Clusters}:  Consider the relatively simple case in the literature (see e.g., \cite{Gesbert, JSDM}) where clusters are separable with non-overlapping AoA ranges. We prove that the \emph{optimal aggregation beamformer has a decomposed architecture} consisting \emph{inner} and \emph{outer components}. The inner components match the dominant eigen-subspaces of different clustered channels to receive low-dimensional signals from them. The outer components then aggregate the weighted signals to compute the desired vector function, where the weights are determined by minimum eigen-values of the said channel eigen-subspaces.

\item{\bf DAB for Overlapping  Clusters}:  Consider the more challenging case of inseparable clusters due to overlapping AoA ranges. We propose an DAB architecture consisting of a single inner and a single outer components. By solving an approximate AirComp-error minimization problem, we prove that the designs of inner and outer DABs can be separated and the separate optimization problems have the identical forms. As a result, the inner DAB performs aggregation over  reduced-dimensional covariance matrices of different clustered channels and the the outer one over small-scale fading channels of different devices.

\item {\bf Clustered-Channel Rank Selection}:  For the case of disjoint clusters, practical algorithms are designed for choosing the ranks of reduced-dimension clustered channels (or received signals) under the criterion of minimum AirComp errors.

\item {\bf Channel Feedback}: To enable the preceding DAB design, schemes are presented for analog channel feedback for both the cases of disjoint and overlapping clusters. The schemes feature simultaneous reduced-dimension feedback   by devices in a same cluster and sequential feedback for different clusters.

\end{itemize}

The paper is organized as follows. In Section II, the system model is  introduced and the  AirComp design problem is formulated. In Section III, the DAB designs are presented for both the cases of disjoint and overlapping channel clusters.  The clustered-channel rank selection problem is solved in Section IV. The analog channel feedback schemes are proposed in Section V. Section VI presents the simulation   results followed by concluding remarks in Section VII.

\section{System Model and Problem Formulation}
\subsection{System Model}
\begin{figure}[t]
\includegraphics[width=\textwidth]{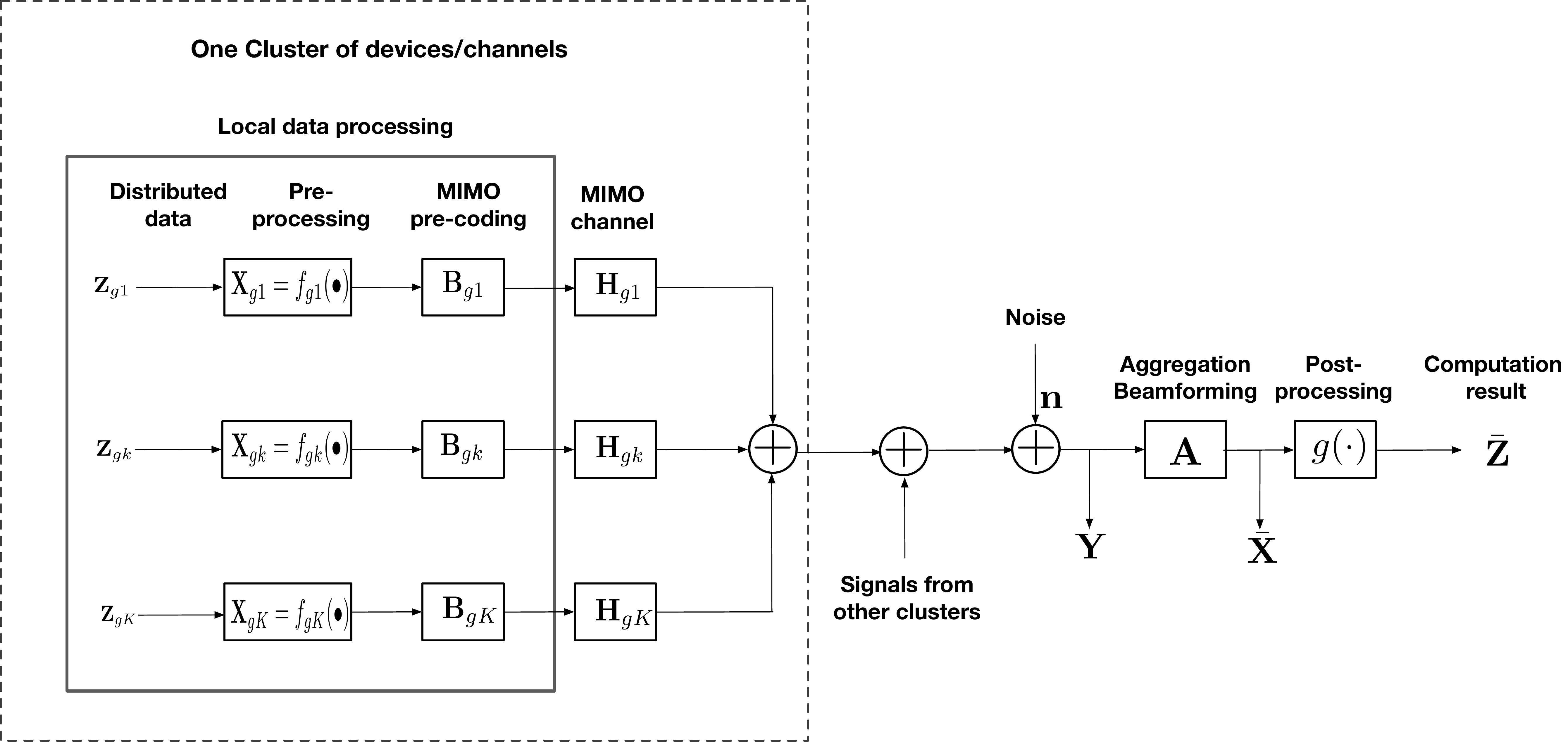}
\caption{Block diagram of MIMO AirComp over a multi-access channel. 
} \label{fig:AirCompModel}
\end{figure}

Consider an IoT system  with one AP and a large number of  edge devices. Perfect local CSI is assumed to be available at all devices and channel reciprocity is considered. The system is designed to perform AirComp of  distributed data transmitted by the devices. The system operations is illustrated in Fig. \ref{fig:AirCompModel} and described as follows. The devices form $G$ clusters each of which comprises $K$ members. The $k$-th device (or channel) in the $g$-th cluster is identified by the indices $(g, k)$.  Each device, say device $(g, k)$,  is provisioned with an array of $N_t$ antennas for transmitting a $L$-dimensional pre-processed  vector symbol by linear analog modulation, which is denoted as ${\bf X}_{g, k}$ representing $f({\bf Z}_{g, k})$ in Fig. \ref{fig:AirCompModel}, to the AP after precoding. The $N_t\times L$ precoding matrix is represented by ${\bf B}_{g, k}$. For simplicity, we assume $N_t=L$, namely exactly $L$ antennas are used to transmit the $L$-dimensional vector symbol. Equipped with an large-scale array of $N_r$ antennas ($N_r\gg N_t$),  the AP receives the simultaneous signals from all devices. The total received signal, denoted as ${\bf Y}$, is given as
\begin{equation}\label{eq:RecSignal}
{\bf Y} = \sum\limits_{g=1}^G\sum\limits_{k=1}^K{\bf Y}_{g,k} =  \sum\limits_{g=1}^G\sum\limits_{k=1}^K{\bf H}_{g, k}{\bf B}_{g, k}{\bf X}_{g, k}+{\bf n},
\end{equation}
where ${\bf Y}_{g,k}$ is the received signal from device $(g,k)$, $\{{\bf H}_{g, k}\}$ represent  the uplink MIMO  channels and ${\bf n}$ is the channel-noise vector comprising $\mathcal{CN}(0,1)$ elements. Then  the received total signal is processed by  aggregation beamforming, represented by the $L\times N_r$ matrix ${\bf A}$,  to yield the desired summation $\sum_{g, k}{\bf X}_{g,k}$, which gives the desired vector-valued function after post-processing (see Fig. \ref{fig:AirCompModel}). The current work focuses on designing ${\bf A}$ to minimize the distortion in functional computation. The distortion is measured by the MSE $\mathsf{E}[||{\bf AY}-\sum_{g, k}{\bf X}_{g, k}||]$, which is the AirComp performance metric throughout the paper.

We adopt the model of clustered MIMO multi-access channels in \cite{JSDM, Gesbert}, characterized by clustered transmitters and rich local scattering. Consequently, for MIMO channels in the same cluster, there exists receive-antenna correlation but no transmit-antenna correlation. Specifically, the spatial correlation of the channels in the $g$-th cluster is represented by the covariance matrix  ${\bf \Psi}_g$ of  rank denoted as $R_g$, namely $\mathsf{E}[{\bf H}_{g, k}{\bf H}_{g, k}^H] = {\bf \Psi}_g$ for given $g$ and any $k$. The rank $R_g$ satisifies the relation $L \leq  R_g \leq N_r$.  The matrix ${\bf \Psi}_g$  is largely determined by \emph{angle-of-arrival} (AoA) range $\Delta\theta_g = [\theta_g, \theta_g']$ as well as array parameters (e.g., topology and antenna spacing). As the AoA ranges of different clusters may be different, $R_g$ may vary in different clusters. Decompose the matrix ${\bf \Psi}_g$ by singular-value decomposition as ${\bf U}_g{\bf \Lambda}_g{\bf U}_g^H$. Then the channel matrix ${\bf H}_{g, k}$ can be written as
\begin{equation}\label{eq:ChModel}
{\bf H}_{g, k} = {\bf U}_g{\bf \Lambda}_g^{\frac{1}{2}}{\bf W}_{g, k},
\end{equation}
where each element of ${\bf W}_{g, k}$ is i.i.d. and follows $\mathcal{CN}(0,1)$.  The array at the AP is assumed to be linear. Consider the case that  the AoA ranges of different clusters are \emph{non-overlapping}, which is referred to as the case of \emph{disjoint  clusters}.
Under this assumption, it is shown  in \cite{JSDM, Gesbert}, as the array size $N_r$ grows, two channels belonging to different clusters approach being orthogonal as a result of ${\bf U}_m^H {\bf U}_n \rightarrow \mathbf{0}$. On the other hand, when the clusters' AoA ranges \emph{overlap}, different clusters of channels cannot be orthogonalized by using a large-scale receive array, which is referred to as the case of \emph{overlapping clusters}. Both cases are considered in the sequel.

\subsection{Problem of Decomposed Aggregation Beamforming}
In this subsection, the AirComp problem for WDA is formulated as a joint DAB matrix, denoising factor, and precoders design problem.

The aggregation beamforming martix is designed under two constraints, namely the constraints of  channel equalization and transmission power, described as follows. To output the desired summation $\sum\nolimits_{g, k}{\bf X}_{g, k}$, the  beamforming martix need be jointly designed with the precoders $\{{\bf B}_{g, k}\}$ to overcome channel distortion. This leads to the \emph{constraints of channel equalization}:
\begin{equation}\label{eq:ChEq}
{\bf A}{\bf H}_{g, k}{\bf B}_{g, k} = \eta {\bf I},~\forall g,k.
\end{equation}
where $\eta$ is a positive scalar, called \emph{denoising factor}. It should be reiterated that the aggreagtion beamforming leverages ``interference" in aggregation instead of suppressing it like  the traditional ZF beamforming, thus requiring much fewer DoF than the latter. Consequently, the ZF constraints in \eqref{eq:ChEq}, indeed for resolving the inter-stream interference, are irrelevant for the beamforming design. Next, each device has finite transmission power, denoted as $P_t$. The power of the pre-processed data symbol ${\bf X}_{g, k}$ is given as $\mathsf{E}\l[\l\|{\bf X}_{g, k}\r\|^2\r]$. Without loss of generality, unit symbol power for all devices is assumed. Then the \emph{transmission-power constraints} can be written as
\begin{equation}\label{Eq:PwrConst}
\tr\big({\bf B}_{g, k}{\bf B}_{g, k}^H\big)\leq P_t,~\forall g,k.
\end{equation}

The objective of designing the beamforming matrix is to minimize the AirComp distortion. The joint design with precoders at devices  can be formulated as an the following optimization problem under the constraints in \eqref{eq:ChEq} and \eqref{Eq:PwrConst}:
\begin{eqnarray*}
~~~~~~~~~~~~~~~~~~~~~~~~~~~~~~~~~~~~
\begin{array}{c}
\min\limits_{{\bf A},\{{\bf B}_{g,k}\}, \eta}\; \mathsf{E}\l[\l\|\dfrac{1}{\eta}{\bf AY}-\sum\limits_{g=1}^G\sum\limits_{k=1}^K{\bf X}_{g, k}\r\|^2\r],
\end{array}~~~~~~~~~~~~~~~~~~~~~~~~\eqref{eq:P1}
\end{eqnarray*}
\vspace{-1.8em}
\begin{subequations}\label{eq:P1}
\begin{align}({\bf P1})\qquad
\text{s.t.}\;\;&{\bf A}{\bf H}_{g, k}{\bf B}_{g, k} = \eta {\bf I},~\forall g,k, \label{eq:P1.1}\\
&\tr\big({\bf B}_{g, k}{\bf B}_{g, k}^H\big)\leq P_t,~\forall g,k. \label{eq:P1.2}
\end{align}
\end{subequations}

\section{Decomposed Aggregation Beamforming}
In this section, Problem ({\bf P1}) is first reduced to an equivalent problem focusing on DAB design. By deriving approximate solution of the non-convex problem, the DAB matrices are designed for both the cases of disjoint and overlapping clusters.

\subsection{An Equivalent DAB Design Problem}
Problem ({\bf P1}) is simplified to a DAB matrix design problem as follows.

First, by substituting \eqref{eq:P1.1}, the objective function of Problem ({\bf P1}) can be rewritten as
\begin{equation}\label{eq:Obj1}
\begin{split}
 & \mathsf{E}\l[ \l\| \dfrac{1}{\eta}{\bf A}\big(\sum\limits_{g=1}^G\sum\limits_{k=1}^K{\bf H}_{g, k}{\bf B}_{g, k}{\bf X}_{g,k}+{\bf n} \big) -\sum\limits_{g=1}^G\sum\limits_{k=1}^K{\bf X}_{g, k} \r\|^2\r],\\
= & \dfrac{1}{\eta^2}\mathsf{E} \l[\l\|{\bf A}{\bf n}\r\|^2\r],\\
= & \dfrac{1}{\eta^2}N_0\tr\big({\bf A}{\bf A}^H\big),
\end{split}
\end{equation}
where $N_0$ is the noise power. One can observe from \eqref{eq:Obj1} that the computation error due to channel noise ${\bf n}$, given by the objective, decreases as $\eta$ grows, giving its name denoising factor.

Next, we derive the optimal $\eta$ and $\{{\bf B}_{g,k}\}$ in terms of ${\bf A}$ based on the constraints \eqref{eq:P1.1} and \eqref{eq:P1.2}. Based on the channel equalization constraints in \eqref{eq:P1.1}, the optimal precoders $\{{\bf B}_{g, k}^*\}$ can be solved as
\begin{equation}\label{eq:Bgk}
{\bf B}^*_{g, k} = \eta ({\bf A}{\bf H}_{g, k})^H({\bf A}{\bf H}_{g, k}{\bf H}_{g, k}^H{\bf A}^H)^{-1},~\forall g,k.
\end{equation}
By substituting ${\bf B}^*_{g, k}$ in \eqref{eq:Bgk} into the transmission-power constraints in \eqref{eq:P1.2}, we have
\begin{equation}
\eta^2\tr\big(({\bf A}{\bf H}_{g, k}{\bf H}_{g, k}^H{\bf A}^H)^{-1}\big)\leq P_t, \forall g,k.
\end{equation}
Equivalently,
\begin{equation}\label{eq:ETACon}
\eta \leq \sqrt{\dfrac{P_t}{\tr\big(({\bf A}{\bf H}_{g, k}{\bf H}_{g, k}^H{\bf A}^H)^{-1}\big)}},~\forall g,k.
\end{equation}
Since the objective function in \eqref{eq:Obj1} decreases with increasing $\eta$, the optimal denoising factor $\eta^*$, constrained by \eqref{eq:ETACon}, is given as
\begin{equation}\label{eq:ETAOpt}
\eta^*=\max \eta =  \min\limits_{g,k}~\sqrt{\dfrac{P_t}{\tr\big(({\bf A}{\bf H}_{g, k}{\bf H}_{g, k}^H{\bf A}^H)^{-1}\big)}}.
\end{equation}
By substituting $\eta^*$ in \eqref{eq:ETAOpt} into \eqref{eq:Bgk}, we can get the optimal precoders, $\{{\bf B}_{g, k}^*\}$. The results are summarized as follows.

\begin{lemma}[Optimal denoising factor and precoding]
Given an aggregation beamforming, ${\bf A}$ the optimal conditional denoising factor and precoders are
\begin{equation}\label{eq:DPOpt}
\boxed{
\begin{aligned}
\bullet \;\; &\text{Optimal denoising factor}: \\
&\eta^* =  \min\limits_{g,k}~\sqrt{\dfrac{P_t}{\tr\big(({\bf A}{\bf H}_{g, k}{\bf H}_{g, k}^H{\bf A}^H)^{-1}\big)}},\\
\bullet \;\; &\text{Optimal precoders}: \\
&{\bf B}_{g, k}^* = \eta^* ({\bf A}{\bf H}_{g, k})^H({\bf A}{\bf H}_{g, k}{\bf H}_{g, k}^H{\bf A}^H)^{-1}, \forall g,k.
\end{aligned}
}
\end{equation}
\end{lemma}
\begin{remark}[Weakest link dominant performance]\emph{
As mentioned, the AirComp error is proportional to $\dfrac{1}{\eta^{*2}}$ and thus it is desirable to enhance the denoising factor $\eta^*$. One can observe from \eqref{eq:DPOpt}, the $\eta^*$ is limited by the weakest link. To be specific, a weak link is characterized by small channel gains, the largest misalignment between the channel matrix and DAB ${\bf A}$, or both. Note that the alignment between a channel, say ${\bf H}_{g,k}$, and ${\bf A}$ can be measured by a sub-space distance \cite{Guangxu}. It follows that the weakest link corresponds to $\max\limits_{g,k}\tr\big(({\bf A}{\bf H}_{g, k}{\bf H}_{g, k}^H{\bf A}^H)^{-1}\big)$ in \eqref{eq:DPOpt}.}
\end{remark}

Last, by substituting the optimal design in \eqref{eq:DPOpt} into \eqref{eq:Obj1}, the unconstrained DAB design problem, equivalent to ({\bf P1}), is derived as
\begin{equation}\label{eq:P2}({\bf P2})\qquad
\min\limits_{{\bf A}\in \mathbb{C}^{L\times N_r}}\max\limits_{g,k}\; \tr\big({\bf A}{\bf A}^H\big) \tr\l(({\bf A}{\bf H}_{g, k}{\bf H}_{g, k}^H{\bf A}^H)^{-1}\r).
\end{equation}
The problem is non-convex. The classic solution approach is \emph{semi-definite relaxation} (SDR), which is, however, too complex in the current context of massive MIMO due to its iterative algorithms and the dimensionality, $N_r\to\infty$. A more efficient approach as we pursue is to exploit high-dimensionality but low rank characteristics of clustered channels to design efficient DAB matrices in closed form. The details are presented in the following sub-sections.

\subsection{DAB Design for Disjoint Clusters}\label{sect:DisDAB}
Consider the case of disjoint clusters where the AoA ranges of any two clusters are disjoint. With large-scale receive arrays ($N_r\to\infty$), it is well known that the column spaces of the covariance matrices of any two differnet cluster channels are orthogonal: ${\bf U}_g^H {\bf U}_{g^{'}} = {\bf 0}, \forall g\not= g^{'}$. Exploiting this property, we first prove that the optimal DAB has a summation form, where each term depends on only one cluster. Furthermore, each summation term is decomposed into a product form cascading an inner and an outer per-cluster beamforming, where the former reduces signal-space dimension and the latter performs AirComp in the reduced-dimensional signal-space.

Given the said orthogonality between cluster channels, the received signals from different clusters of devices can be decoupled without inter-cluster interference. This fact allows us to derive the structure of the optimal DAB as shown below.

\begin{proposition}[Decomposed DAB structure]\label{lma:Decompose}
In the case of disjoint clusters, the optimal DAB matrix solves Problem ({\bf P2}) has the following decomposed structure,
\begin{equation}\label{eq:Decompose}
{\bf A}^* = \sum\limits_{g=1}^G {\bf C}_g^H  {\bf U}_g^H,
\end{equation}
where the size of ${\bf C}_g$ is $R_g\times L$.
\end{proposition}
\proof See Appendix \ref{apdx:lmaDCP}.

Several observations can be made from the optimal DAB structure in \eqref{eq:Decompose}. Let $\{{\bf U}_g\}$ and $\{{\bf C}_g\}$ be referred to as the inner and the outer per-cluster DABs, respectively. Each inner term, ${\bf U}_g$, is matched to one cluster and extracts the signal from the dominant $R_g$-dimensional eigen-space of the high-dimensional cluster channel, $\left\{{\bf H}_{g,k},k\in[1,K]\right\}$. This yields a reduced-dimensional signal-space, where performing AirComp using $\{{\bf C}_g\}$ has two advantages. The SNR therein are high and the sub-space distances between the effective channels $\left\{{\bf U}_g^H{\bf H}_{g,k},k\in[1,K]\right\}$ are small, leading to AirComp-error reduction. Furthermore, AirComp in a reduced-dimensional space results in dramatic complexity reduction.

Next, building on the optimal DAB structure on \eqref{eq:Decompose}, we focus on designing the outer per-cluster DABs $\{{\bf C}_g\}$. By substituting \eqref{eq:Decompose}, Problem ({\bf P2}) can be derived as
\begin{equation}\label{eq:P3}({\bf P3})\qquad
\min\limits_{\{{\bf C}_g\}} \max\limits_{g,k}\;  \sum\limits_{m=1}^G \tr\l( {\bf C}_m^H {\bf C}_m \r) \sum\limits_{i=1}^L \dfrac{1}{ \lambda_i\l( {\bf C}_g^H{\bf F}_{g,k} {\bf F}_{g,k}^H {\bf C}_g \r)},
\end{equation}
where ${\bf F}_{g,k} = {\bf \Lambda}_g^{\frac{1}{2}}{\bf W}_{g,k}$ is the effective channel after dimension reduction, the function $\lambda_i\l( \cdot \r)$ acquires the $i$-th eigenvalue of a matrix, and the eigenvalues are arranged in a decreasing order, i.e., $\lambda_1\l(\cdot\r)\geq \lambda_2\l(\cdot\r)\geq...\geq\lambda_L\l(\cdot\r)$.

Problem ({\bf P3}) reduces high-dimensional design in Problem ({\bf P2}) to the design of reduced-dimensional DAB matrices $\{{\bf C}_g\}$. Problem ({\bf P3}) is non-convex. As its solution is intractable, we derive an approximate solution in closed form to obtain an efficient design of $\{{\bf C}_g\}$. The approximation consists of two steps. The first is to replace the objective function in Problem $({\bf P3})$ by an upper bound based on the following inequalities, $\lambda_i\l( {\bf C}_g^H{\bf F}_{g,k} {\bf F}_{g,k}^H {\bf C}_g \r)\geq \lambda_{\min}\l( {\bf C}_g^H{\bf F}_{g,k} {\bf F}_{g,k}^H {\bf C}_g \r), \forall i$. The bounds are tight when the eigenvalues of the matrix $\l( {\bf C}_g^H{\bf F}_{g,k} {\bf F}_{g,k}^H {\bf C}_g \r)$ are similar. It follows that Problem ({\bf P3}) can be approximated as
\begin{equation}\label{eq:P4}({\bf P4})\qquad
\min\limits_{\{{\bf C}_g\}} \max\limits_{g,k}\; \sum\limits_{m=1}^G \tr\l( {\bf C}_m^H {\bf C}_m \r) \lambda_{\min}^{-1}\l( {\bf C}_g^H{\bf F}_{g,k} {\bf F}_{g,k}^H {\bf C}_g \r)L,
\end{equation}
However, Problem ({\bf P4}) is still non-convex. To overcome the difficulty, the second approximation step adopts a general approach in beamforming literatures (see e.g., \cite{UniBF1,UniBF2,UniBF3}), that constrain the beamforming matrices $\{{\bf C}_g\}$ to be unitary. This is reasonable as it is the sub-space spanned by ${\bf C}_g$ that has a dominant effect on the AirComp performance. With the constraint (${\bf C}_g^H{\bf C}_g={\bf I}$), Problem ({\bf P4}) can be further approximated as
\begin{equation}({\bf P5})\qquad
\begin{aligned}
\mathop{\min }\limits_{\{{\bf C}_g\}} \max\limits_{g,k}\;  &\lambda_{\min}^{-1}\l( {\bf C}_g^H{\bf F}_{g,k} {\bf F}_{g,k}^H {\bf C}_g \r), \\
{\textmd{s.t.}}\;\; &{\bf C}_g^H{\bf C}_g={\bf I},~\forall g.
\end{aligned}
\end{equation}
In Problem ({\bf P5}), it can be shown that the design of ${\bf C}_g$ depends on solely the $g$-th cluster channel matrices $\{{\bf F}_{g,k},1\leq k\leq K\}$ and is independent of other clusters ($g^{'}\not=g$). Therefore, the inner beamforming design can be decoupled, as shown in the following lemma.
\begin{lemma}[Outer per-cluster DAB]\label{lma:Outer}
The joint design of $\{{\bf C}_g\}$ in Problem {\bf P5} can be decoupled to solve ${\bf C}_g$ in the following problem for all $g$.
\begin{equation}({\bf P6})\qquad
\begin{aligned}
\mathop{\min }\limits_{{\bf C}_g} \max\limits_{k}\;  &\lambda_{\min}^{-1}\l( {\bf C}_g^H{\bf F}_{g,k} {\bf F}_{g,k}^H {\bf C}_g \r), \\
{\textmd{s.t.}}\;\; &{\bf C}_g^H{\bf C}_g={\bf I}.
\end{aligned}
\end{equation}
\end{lemma}

Problem ({\bf P6}) has the same form as the problem ({\bf P5}) in \cite{Guangxu}. Following the approach in \cite{Guangxu}, ${\bf C}^*_g$, that solves Problem ({\bf P6}), can be obtained as the weighted sub-space centroid of the column spaces of $\l\{ {\bf F}_{g,k} \r\}$, which is the $L$-dimensional principal eigen-space of the following matrix,
\begin{equation}\label{eq:Sga}
{\bf S}_g^{({\rm a})} = \sum\limits_{k=1}^K\lambda_{\min}\big({\bf F}_{g, k}^H{\bf F}_{g, k}\big){\bf U}_{{\bf F}_{g, k}}{\bf U}_{{\bf F}_{g, k}}^H,
\end{equation}
where ${\bf U}_{{\bf F}_{g, k}}$ is the column space of ${\bf F}_{g, k}$. In other words, the solution is ${\bf C}^*_g = \l[{\bf U}_{{\bf S}_g^{({\rm a})}}\r]_{1:L}$, where $\l[{\bf U}_{{\bf S}_g^{({\rm a})}}\r]_{1:L}$ denotes the $L$-dimensional principal eigen-space of ${\bf S}_g^{({\rm a})}$.

By combining the results in Lemma \ref{lma:Decompose} and \ref{lma:Outer}, the DAB matrix design for the AirComp in disjoint-cluster case is given as
\begin{equation}\label{eq:DAB}
\text{(Optimal DAB)}\qquad{\bf A}^* = \sum\limits_{g=1}^G {\bf C}_g^{*H} {\bf U}_g^H,
\end{equation}
where ${\bf C}^*_g = \l[{\bf U}_{{\bf S}_g^{({\rm a})}}\r]_{1:L}$ and ${\bf S}_g^{({\rm a})}$ is defined in \eqref{eq:Sga}, respectively.

\subsection{DAB Design for Overlapping Clusters}\label{sect:OvpDAB}
In this subsection, DAB is designed for the case of overlapping clusters. Unlike the preceding case of disjoint clusters, it is impossible to decouple the received signals from different clusters due to their overlapping \cite{JSDM}. Then, the optimal DAB form in \eqref{eq:DAB} derived for the former not longer for the current case. Nevertheless, inspired by the result, we propose that the DAB design should have the decomposed form: ${\bf A}={\bf A}_{({\rm o})}{\bf A}_{({\rm i})}$, where the inner DAB ${\bf A}_{({\rm i})}$ is a $R_{\rm s}\times N_r$ matrix with
\begin{equation}\label{eq:Rs}
R_{\rm s}=\min\l(R_1,...,R_g,...,R_G\r),
\end{equation}
and the outer DAB ${\bf A}_{({\rm o})}$ is a $L\times R_{\rm s}$ matrix. In other words, ${\bf A}_{({\rm i})}$ is responsible for the dimension reduction of the signal space. Nevertheless, the operation of AirComp is distributed over outer and inner beamformers instead of relying only on the former as in the preceding case. For tractability, following the same reason as for designing ${\bf C}_g$ in Problem ({\bf P4}), we constrain both ${\bf A}_{({\rm o})}^H$ and ${\bf A}_{({\rm i})}^H$ to be unitary. We show in the sequel that the designs of outer and inner DAB can be reduced to optimization problems having the identical form.

\subsubsection{Inner beamforming design}
Under the criterion of minimizing AirComp error, the inner DAB should be matched to the $R_{\rm s}$-dimensional dominant eigen-space of each cluster channel, which is obtained as follows.

Denote the $R_{\rm s}$-dimensional dominant eigen-space, the $R_{\rm s}$-dimensional dominant eigenvalue matrix, and the corresponding small-scale fading matrix of device $(g,k)$ as ${\bf \hat{U}}_g = \l[{\bf U}_g\r]_{1:R_{\rm s}}$, ${\bf \hat{\Lambda}}_g=\l[{\bf \Lambda}_g\r]_{1:R_{\rm s},1:R_{\rm s}}$, and ${\bf \hat{W}}_{g,k} = \l[{\bf W}_{g,k}\r]_{1:R_{\rm s},:}$,
respectively. Then, the dominant $R_{\rm s}$-dimensional sub-space of the channel model in \eqref{eq:ChModel} is
\begin{equation}\label{eq:ExModel}
{\bf \hat{H}}_{g,k} = {\bf \hat{U}}_g{\bf \hat{\Lambda}}_g^{\frac{1}{2}}{\bf \hat{W}}_{g,k}.
\end{equation}
To solve Problem ({\bf P2}) in this case, we first derive a useful inequality as follows.
\begin{lemma}\label{lma:EGA}
With ${\bf A}={\bf A}_{({\rm o})}{\bf A}_{({\rm i})}$, the following inequality holds.
\begin{equation}\label{eq:EGA}
 \tr\l(({\bf A}{\bf \hat{H}}_{g, k}{\bf \hat{H}}_{g, k}^H{\bf A}^H)^{-1}\r) \leq \lambda_{\min}^{-1} \l({\bf A}_{({\rm i})}{\bf \hat{U}}_g {\bf \hat{U}}_g^H{\bf A}_{({\rm i})}^H \r)\sum\limits_{i=1}^{L} \lambda_i^{-1}\big( {\bf \hat{\Lambda}}_g^{\frac{1}{2}} {\bf \hat{W}}_{g,k}{\bf \hat{W}}_{g,k}^H {\bf \hat{\Lambda}}_g^{\frac{1}{2}} \big).
\end{equation}
\end{lemma}
\proof See Appendix \ref{apdx:lmaEGA}.

By substituting \eqref{eq:EGA}, Problem ({\bf P2}) can be approximated as
\begin{equation}({\bf P7})\qquad
\begin{aligned}
\mathop{\min }\limits_{{\bf A}_{({\rm i})}} \max\limits_{g}\;  &\alpha^{'}_{g}\lambda^{-1}_{\min}\big({\bf A}_{({\rm i})}{\bf \hat{U}}_g{\bf \hat{U}}_g^H{\bf A}_{({\rm i})}^H\big), \\
{\textmd{s.t.}}\;\; &{\bf A}_{({\rm i})}{\bf A}_{({\rm i})}^H={\bf I},
\end{aligned}
\end{equation}
where $\alpha^{'}_{g} = \max\nolimits_k \sum\nolimits_{i=1}^{L} \lambda_i^{-1}\big( {\bf \hat{\Lambda}}_g^{\frac{1}{2}} {\bf \hat{W}}_{g,k}{\bf \hat{W}}_{g,k}^H {\bf \hat{\Lambda}}_g^{\frac{1}{2}} \big)$.
The problem has the same form as Problem ({\bf P5}) in \cite{Guangxu}. Following the approach in \cite{Guangxu},  ${\bf A}_{({\rm i})}^H$ is solved as the $R_{\rm s}$-dimensional principal eigen-space of the following matrix,
\begin{equation}\label{eq:Sb}
{\bf S}^{({\rm b})} = \sum\limits_{g=1}^G \alpha^{'}_{g} {\bf \hat{U}}_g{\bf \hat{U}}_g^H.
\end{equation}
That's to say, the solution is ${\bf A}^*_{({\rm i})} = \l[{\bf U}_{{\bf S}^{({\rm b})}}\r]_{1:R_{\rm s}}^H$, where $\l[{\bf U}_{{\bf S}^{({\rm b})}}\r]_{1:R_{\rm s}}$ is the $R_{\rm s}$-dimensional principal eigen-space of ${\bf S}^{({\rm b})}$.

\subsubsection{Outer beamforming design}
By substituting ${\bf A}^*_{({\rm i})} = \l[{\bf U}_{{\bf S}^{({\rm b})}}\r]_{1:R_{\rm s}}^H$ into Problem ({\bf P2}), it can be derived as
\begin{equation}({\bf P8})\qquad
\begin{aligned}
\mathop{\min }\limits_{{\bf A}_{({\rm o})}} \max\limits_{g,k}\;  &\tr\l(\l({\bf A}_{({\rm o})} {\bf F}_{g,k} {\bf F}_{g,k}^H {\bf A}_{({\rm o})}^H\r)^{-1}\r), \\
{\textmd{s.t.}}\;\; &{\bf A}_{({\rm o})}{\bf A}_{({\rm o})}^H={\bf I},
\end{aligned}
\end{equation}
where ${\bf F}_{g,k} = {\bf A}^*_{({\rm i})} {\bf \hat{U}}_g{\bf \hat{\Lambda}}_g^{\frac{1}{2}}{\bf \hat{W}}_{g,k}$. Using the following inequality,
\begin{equation}
\tr\l(\l({\bf A}_{({\rm o})} {\bf F}_{g,k} {\bf F}_{g,k}^H {\bf A}_{({\rm o})}^H\r)^{-1}\r) \leq L\lambda^{-1}_{\min}\l({\bf A}_{({\rm o})} {\bf F}_{g,k} {\bf F}_{g,k}^H {\bf A}_{({\rm o})}^H\r),
\end{equation}
Problem ({\bf P8}) can be further approximated to
\begin{equation}({\bf P9})\qquad
\begin{aligned}
\mathop{\min }\limits_{{\bf A}_{({\rm o})}} \max\limits_{g,k}\;  &L\lambda^{-1}_{\min}\l({\bf A}_{({\rm o})} {\bf F}_{g,k} {\bf F}_{g,k}^H {\bf A}_{({\rm o})}^H\r), \\
{\textmd{s.t.}}\;\; &{\bf A}_{({\rm o})}{\bf A}_{({\rm o})}^H={\bf I}.
\end{aligned}
\end{equation}
Then ${\bf A}_{({\rm o})}^H$ can be solved by the same approach with Problem ({\bf P6}) as the $L$-dimensional principal eigen-space of the following matrix,
\begin{equation}\label{eq:Sc}
{\bf S}^{({\rm c})} = \sum\limits_{g=1}^G\sum\limits_{k=1}^K  \lambda_{\min}\l({\bf F}_{g,k}^H{\bf F}_{g,k} \r) {\bf U}_{{\bf F}_{g,k}} {\bf U}_{{\bf F}_{g,k}}^H.
\end{equation}
That's to say, the solution is ${\bf A}_{({\rm o})}^* = \l[{\bf U}_{{\bf S}^{({\rm c})}}\r]_{1:L}^H$, where $\l[{\bf U}_{{\bf S}^{({\rm c})}}\r]_{1:L}$ is the $L$-dimensional principal eigen-space of ${\bf S}^{({\rm c})}$.

\subsubsection{Overall DAB design}
In summary, the overall DAB design in overlapping-cluster case is comprised of inner beamformer ${\bf A}_{({\rm i})}^*$ and outer beamformer ${\bf A}_{({\rm o})}^*$, which are given as
\begin{equation}\label{eq:CorAB}
\boxed{
{\bf A}^*_{({\rm i})} = \l[{\bf U}_{{\bf S}^{({\rm b})}}\r]_{1:R_{\rm s}}^H,\quad {\bf A}_{({\rm o})}^* = \l[ {\bf U}_{{\bf S}^{({\rm c})}}\r]_{1:L}^H,
}
\end{equation}
where ${\bf U}_{{\bf S}^{({\rm b})}}$ and ${\bf U}_{{\bf S}^{({\rm c})}}$ are the $R_{\rm s}$-dimensional and $L$-dimensional principal eigen-space of ${\bf S}^{({\rm b})}$ and ${\bf S}^{({\rm c})}$, and  ${\bf S}^{({\rm b})}$ and ${\bf S}^{({\rm c})}$ are defined in \eqref{eq:Sb} and \eqref{eq:Sc}, respectively.

\begin{remark}[DAB design for overlapping clusters]\emph{
One can observe from \eqref{eq:CorAB} that the DAB design performs two-tier AirComp. To be specific, the inner DAB ${\bf A}_{({\rm i})}$ performs AirComp over channel covariance matrices. Subsequentially, in the reduced-dimension signal space created by the inner DAB, the outer DAB ${\bf A}_{({\rm o})}^*$ performs AirComp over small scale-fading channels.}
\end{remark}

\section{Clustered-Channel Rank Selection}
In the preceding section, DAB is designed with fixed ranks for its components. Adjusting the ranks according to clustered channel covariance provides another dimension for reducing AirComp error. Relevant algorithms are presented in this section.

\subsection{Channel-Rank Selection for Disjoint Clusters}
Modifying the DAB design in \eqref{eq:DAB} to allow variable ranks for inner components:
\begin{equation}\label{eq:DisRank}
{\bf A}^* = \sum\limits_{g=1}^G {\bf \hat{C}}_g^{*H} {\bf \hat{U}}_g^{H},
\end{equation}
where ${\bf \hat{U}}_g^{H}$ selects the $r_g$-dimensional dominant eigen-space of the channel covariance matrix ${\bf \Psi}_g$, and ${\bf \hat{C}}_g^{*}$ is computed in the same way as ${\bf C}_g^*$ with ${\bf U}_g$ replaced by ${\bf \hat{U}}_g$. There exists a tradeoff in setting the ranks of $\{{\bf U}_g\}$. On one hand, as can be proved, increasing the ranks $\{r_g\}$ receives more signal energy from the channels and helps reduce AirComp error. On the other hand, increasing an inner-DAB rank, says $r_g$, increases the dimensionality of the reduced-dimension sub-space, where small-scale-fading channels of cluster $g$ are jointly equalized for the purpose of AirComp, thereby increasing its error. The above tradeoff is leveraged in the sequel to formulate an optimization problem for channel-rank selection and to derive an algorithmic solution.

The problem of channel-rank selection, namely optimizing the ranks $\{r_g\}$ of inner DAB can be formulated by substituting the optimal design in \eqref{eq:DAB} into Problem ({\bf P6}):
\begin{equation}({\bf P10})\qquad
\begin{aligned}
\mathop{\min }\limits_{\{r_g\}} \max\limits_{g,k}\;  &\lambda_{\min}^{-1}\l({\bf C}_g^{*H}{\bf F}_{g, k}{\bf F}_{g, k}^H{\bf C}_g^*\r), \\
{\textmd{s.t.}}\;\; &L\leq r_g\leq R_g,~\forall g.
\end{aligned}
\end{equation}
where ${\bf F}_{g, k} = {\bf \Lambda}_g^{\frac{1}{2}}{\bf W}_{g,k}$ is the $r_g\times L$ effective channel after dimension reduction using inner DAB. Decompose ${\bf F}_{g, k}$ using SVD as ${\bf F}_{g, k} = {\bf U}_{{\bf F}_{g, k}} {\bf \Sigma}_{{\bf F}_{g,k}} {\bf V}_{{\bf F}_{g, k}}^H$. Then, the objective function of Problem ({\bf P10}) can be bounded as
\begin{equation}\label{eq:FgkRelax}
\begin{split}
\lambda_{\min}^{-1}\l({\bf C}_g^{*H}{\bf F}_{g, k}{\bf F}_{g, k}^H{\bf C}_g^*\r) &= \lambda_{\min}^{-1}\l({\bf C}_g^{*H}{\bf U}_{{\bf F}_{g, k}} {\bf \Sigma}_{{\bf F}_{g,k}}^2 {\bf U}_{{\bf F}_{g, k}}^H{\bf C}_g^*\r),\\
&\leq \lambda_{\min}^{-1} \l({\bf \Sigma}_{{\bf F}_{g,k}}^2\r) \lambda_{\min}^{-1}\l({\bf C}_g^{*H}{\bf U}_{{\bf F}_{g, k}}{\bf U}_{{\bf F}_{g, k}}^H{\bf C}_g^*\r), \\
& = \lambda_{\min}^{-1} \l({\bf F}_{g, k}^H{\bf F}_{g, k}\r) \l( 1-{\rm d}_{\rm P2}^2\l( {\bf C}_g^{* },  {\bf U}_{{\bf F}_{g, k}} \r)\r)^{-1},
\end{split}
\end{equation}
where ${\rm d}_{\rm P2}^2\l( {\bf C}_g^{*},  {\bf U}_{{\bf F}_{g, k}} \r)$ is the projection 2-norm sub-space distance between the sub-spaces spanned by ${\bf C}_g^{*}$ and ${\bf U}_{{\bf F}_{g, k}}$ \cite{GRSM_SubSp}. Using the inequality in \eqref{eq:FgkRelax}, Problem ({\bf P10}) can be further approximated for tractability as
\begin{equation}({\bf P11})\qquad
\begin{aligned}
\mathop{\min }\limits_{\{r_g\}} \max\limits_{g,k}\;  &\lambda_{\min}^{-1} \l( {\bf F}_{g, k}^H {\bf F}_{g, k}\r) \l( 1-{\rm d}_{\rm P2}^2\l( {\bf C}_g^{*},  {\bf U}_{{\bf F}_{g, k}} \r)\r)^{-1}, \\
{\textmd{s.t.}}\;\; &L\leq r_g\leq R_g,~\forall g.
\end{aligned}
\end{equation}
The objective function in Problem ({\bf P11}) represents a component of AirComp error measured using MSE. A useful result is obtained as follows.
\begin{lemma}\label{lma:EGVRank}
Consider the $r_g\times N_t$ effective channel ${\bf F}_{g,k}$ of device $k$ in cluster $g$ after dimension reduction. The eigenvalue $\lambda_{\min} \l( {\bf F}_{g, k}^H {\bf F}_{g, k}\r)$ is a monotone increasing function of $r_g$.

\end{lemma}

\proof See Appendix \ref{apdx:lmaEGVRank}.

\begin{remark}[Tradeoff in channel-rank selection]\emph{
The said tradeoff is reflected in the objective function in Problme ({\bf P11}). To be specific, as $r_g$ grows, $\lambda_{\min} \l( {\bf F}_{g, k}^H {\bf F}_{g, k}\r)$ increases according to Lemma \ref{lma:EGVRank}, reducing AirComp error. On the other hand, the dimensionality ($r_g$) of the sub-space of ${\bf H}_{g,k}$ after dimension reduction grows. Note that in this sub-space, the outer DAB ${\bf \hat{C}}_g^*$ equalizes the cluster of channels $\{{\bf F}_{g,k}, 1\leq k\leq K\}$ for the purpose of AirComp. As the dimensionality grows, the sub-spaces distance, ${\rm d}_{\rm P2}^2\l( {\bf C}_g^{* },  {\bf U}_{{\bf F}_{g, k}} \r)$, increases \cite{GRSM_SubSp}, thereby elevating the AirComp error.}
\end{remark}

For notation simplicity, define $\mathsf{MSE}_{g,k} = \lambda_{\min}^{-1} \l( {\bf F}_{g, k}^H {\bf F}_{g, k}\r) \l( 1-{\rm d}_{\rm P2}^2\l( {\bf C}_g^{*},  {\bf U}_{{\bf F}_{g, k}} \r)\r)^{-1}$
and $\mathsf{MSE}=\max_{g,k} \mathsf{MSE}_{g,k}$. Hence,  Problem ({\bf P11}) can be simplified as
\begin{equation}({\bf P12})\qquad
\begin{aligned}
\mathop{\min }\limits_{r_g}\;  &\mathsf{MSE}, \\
{\textmd{s.t.}}\;\; &L\leq r_g\leq R_g,~\forall g.
\end{aligned}
\end{equation}
In the sequel, Problem ({\bf P12}) is solved to yield two schemes: homogeneous and heterogeneous channel-rank selection.

\subsubsection{Homogeneous rank-selection scheme}
To simplify design, apply the constraint of homogeneous rank selection: $r_g=r,~\forall g$. Then, Problem ({\bf P12}) can be re-written as
\begin{equation}({\bf P13})\qquad
\begin{aligned}
\mathop{\min }\limits_{r}\;  &\mathsf{MSE}, \\
{\textmd{s.t.}}\;\; &r_g=r,~\forall g,\\
 &L\leq r\leq \min\limits_g\{R_g\}.
\end{aligned}
\end{equation}
Since $r$ is an integer variable and its range, $N_t\leq r\leq \min\nolimits_g \{R_g\}$, is usually small, the optimal value of $r$ can be found by one-dimensional search.

\subsubsection{Heterogenous rank-selection scheme}
In this case, inner DAB components $\{{\bf U}_g\}$ are allowed to have different ranks. The corresponding Problem ({\bf P12}) is an integer problem, whose solution is NP-hard. To address this issue, we propose a sub-optimal design based on the following procedure.

First, the channel cluster that is the bottleneck of AirComp is identified and the rank of the corresponding outer DAB component is optimized. Next, the preceding step is repeated till the algorithm converges. The details of the algorithm are in Algorithm \ref{Ag:1}.

\begin{algorithm}
\caption{Heterogeneous rank selection algorithm}\label{Ag:1}
1: Initialize $r_g=N_t, \forall g$.

2: {\bf{Loop}}

3: ~~~~Find $(G_0,K_0)=\arg\max\limits_{g,k} \mathsf{MSE}_{g, k}$,

4: ~~~~Solve the the following problem and find the optimal rank for cluster $G_0$ with fixed $r_g, \forall g\not=G_0$.
\begin{equation*}
\tilde{R}=\arg\min\limits_{r_{G_0}}\max\limits_{k} \mathsf{MSE}_{G_0, k},\quad {\rm s.t.}~ r_{G_0}\leq R_{G_0},
\end{equation*}

5: ~~~~Update $r_{G_0}=\tilde{R}$.

6: {\bf{Until}} convergence.
\end{algorithm}

\subsection{Channel-Rank Selection for Overlapping  Clusters}
Due to overlapping clusters, it is no longer feasible to match the ranks of individual DAB components according to those of individual clusters. However, it is possible to optimize the rank of inner DAB $r$ in the design in \eqref{eq:CorAB} over the range $L \leq r \leq \min\nolimits_g R_g$, namely performing \emph{homogeneous rank selection} similarly as in Problem ({\bf P13}). The resultant  problem of channel-rank selection for  outer DAB can be formulated by substituting the design in \eqref{eq:CorAB} into Problem ({\bf P9}):
\begin{equation}({\bf P14})\qquad
\begin{aligned}
\mathop{\min }\limits_{r}\;  &\lambda^{-1}_{\min}\l({\bf A}_{({\rm o})}^{*} {\bf F}_{g,k} {\bf F}_{g,k}^H {\bf A}_{({\rm o})}^{*H}\r), \\
{\textmd{s.t.}}\;\; &r_g=r,~\forall g,\\
 &L\leq r\leq \min\limits_g\{R_g\},
\end{aligned}
\end{equation}
which can also be solved by one-dimensional search, since the ranks' range, $L\leq r\leq \min\nolimits_g R_g$, is usually small.

\section{Analog Channel Feedback}
In this section, the principle of AirComp is applied to design efficient scheme for CSI feedback to enable DAB designed in the preceding sections. Specifically, given channel reciprocity and reliable feedback channel, the schemes feature low-latency simultaneous analog feedback such that the desired DAB ${\bf A}^*$ can be computed as
\begin{equation}\label{eq:FDBKModel}
{\bf A}^* = q\left( \sum_{g=1}^Gq_g\left({\bf Y}_g\right) \right) = q\left(\sum_{g=1}^G q_g\left(\sum_{k=1}^K {\bf H}_{g,k} {\bf Z}_{g,k} \right)\right),
\end{equation}
where ${\bf Y}_g$ is the aggregated feedback signals from all devices in cluster $g$, $q_g(\cdot)$ is the cluster-based post-processing function, and $q(\cdot)$ is the overall post-processing function. The principle was first applied in \cite{Guangxu} to design  feedback for AirComp targeting rich-scattering channels. Based on the same principle, we design feedback schemes for reduced-dimensional AirComp for clustered MIMO channels.

In practical systems such as 3GPP LTE, CSI feedback is part of control signalling and protected against channel fading and  noise by high transmission power and coding, creating reliable feedback channels. Such channels are also assumed in this work, where reliable analog feedback is ensured by high power and linear analog coding. As a result, noise as well as analog feedback detection \cite{Marzetta} are omitted in the exposition for brevity. In the sequel, we focus on the design of feedback signals, pre-processing, and post-processing.

\subsection{Analog Feedback for Disjoint Clusters}
Based on the design in \eqref{eq:DAB}, the  objective for feedback  is to obtain at the AP  the desired DAB  ${\bf A}^*=\sum\nolimits_{g=1}^G {\bf C}_g^{*H}{\bf U}_g^H$. The matrix ${\bf U}_g$ is the eigen-space of the channel covariance matrix ${\bf \Psi}_g$, which can be estimated reliably at both the AP and devices from past transmission \cite{Gesbert,JSDM}. It follows that the feedback purpose is for the AP to acquire
$\{{\bf C}_g^*\}$, which depend on small-scale fading.

Based on the principle in \eqref{eq:FDBKModel}, we propose the following ``one-shot" analog feedback scheme, where the notation follows that in Section \ref{sect:DisDAB}.
\begin{equation}\nonumber
\boxed{
\begin{aligned}
&\qquad \qquad \qquad\text{{\bf Analog Feedback for Disjoint Clusters}} \\
\bullet \;\; &\text{Individual feedback signals}: {\bf Z}_{g,k} =  \lambda_{\min}\l({\bf F}_{g, k}^H{\bf F}_{g, k}\r) {\bf V}_{{\bf F}_{g, k}}{\bf \Sigma}_{{\bf F}_{g, k}}^{-1}{\bf U}_{{\bf F}_{g, k}}^H, \ \forall g,k,\\
\bullet \;\; &\text{Received/aggregated feedback signal}:  ~~{\bf Y} = \sum\nolimits_{g=1}^G{\bf U}_g \sum\nolimits_{k=1}^K  {\bf F}_{g, k} {\bf Z}_{g, k}, \\
\bullet \;\; &\text{Cluster-based post-processing $q_g(\cdot)$}: ~{\bf Y}_g = {\bf U}_g^H{\bf Y},\\
\bullet \;\; &\text{Overall post-processing $q(\cdot)$}: ~{\bf A}^* = \sum\nolimits_{g=1}^G\l[{\bf U}_{{\bf Y}_g}\r]_{1:L} ^{H}{\bf U}_g^H,
\end{aligned}
}
\end{equation}
where $\l[{\bf U}_{{\bf Y}_g}\r]_{1:L}$ is the $L$-dimensional principal eigen-space of ${\bf Y}_g$. It is straightforward to verify that the DAB obtained using the above feedback scheme is the desired one.

\begin{remark}[One-shot feedback] \emph{The key feature of the above analog feedback scheme is \emph{simultaneous analog transmission} (or "one-shot" feedback). The minimum feedback duration is a single symbol duration. Therefore, the feedback overhead is low and its latency is independent of the number of devices.}
\end{remark}

\subsection{Analog  Feedback for Overlapping  Clusters}

Following the design in Section \ref{sect:OvpDAB}, the desired  DAB in the current case  is ${\bf A}^* = {\bf A}_{({\rm o})}^*{\bf A}_{({\rm i})}^*$. In the preceding case of disjoint clusters, one-shot feedback is feasible due to the fact that signals from different clusters are separable at the AP. This does not hold in the current case while feedback signals from different clusters still need be separated. Consequently, the feedback scheme requires $G$ slots where feedback in  each slot  targets  one specific cluster of  channels.

Consider feedback of the inner DAB ${\bf A}^*_{({\rm i})}$.  One can observe from \eqref{eq:Sb} that the inner DAB ${\bf A}^*_{({\rm i})}$ depends on 1) the channel covariance matrices, which are known to both the AP and devices, and 2) aggregation weights (scalars) $\{\alpha'_g\}$ with one for each cluster. A particular weight, say $\alpha'_g$, requires computation of the maximum over $K$ scalars transmitted by $K$ devices in cluster $g$. This can be realized using the existing AirComp algorithm in \cite{aircomp2}, referred to as \emph{maximum-AirComp algorithm}. As the scalars depend on small-scale fading, their feedback need be periodic and repeated for every channel coherence time. Next, consider feedback of the outer DAB ${\bf A}^*_{({\rm o})}$. The design depends on small-scale fading according to  \eqref{eq:Sc}. For the reason mentioned earlier, the outer-DAB feedback requires $G$ slots.

The proposed scheme combining feedback of outer and inner DAB are shown below, where the notation follows that in Section \ref{sect:OvpDAB}.
\begin{equation}\nonumber
\boxed{
\begin{aligned}
& \qquad \qquad \qquad \quad \text{\bf Analog Feedback for Overlapping Clusters} \\
\bullet& \;\;\text{For }g = 1, 2, \cdots, G,\\
& \;\;\;\;\bullet \;\; \text{Feedback of $\alpha'_g$ using the maximum-AirComp algorithm in \cite{Guangxu}}, \\
& \;\;\;\;\bullet \;\; \text{Individual feedback signals}: {\bf Z}_{g,k}^* =  \lambda_{\min}\l({\bf F}_{g,k}^H{\bf F}_{g,k}\r) {\bf V}_{{\bf F}_{g,k}}{\bf \Sigma}_{{\bf F}_{g,k}}^{-1}{\bf U}_{{\bf F}_{g,k}}^H, \\
& \;\;\;\;\bullet \;\; \text{Received/aggregative  signal from cluster $g$}: {\bf Y}_g =  \sum\nolimits_{k=1}^K  {\bf U}_g {\bf \Lambda}_g^{\frac{1}{2}}{\bf W}_{g,k} {\bf Z}^*_{g,k}, \\
& \;\;\;\;\bullet \;\; \text{Cluster-based post-processing} : ~{\bf \Omega}_g = {\bf A}_{({\rm i})}^*{\bf \hat{U}}_g{\bf \hat{U}}_g^H {\bf Y}_g,\\
\bullet& \;\;\text{End }\\
\bullet& \;\;\text{Compute }{\bf S}^{({\rm b})} = \sum\nolimits_{g=1}^G \alpha^{'}_{g} {\bf \hat{U}}_g{\bf \hat{U}}_g^H, \\
\bullet& \;\;\text{Compute the inner DAB as: } {\bf A}^*_{({\rm i})} = \l[{\bf U}_{{\bf S}^{({\rm b})}}\r]_{1:R_{\rm s}}^H, \\
\bullet& \;\;\text{Compute }{\bf S}^{({\rm c})} = \left[\sum\nolimits_{g=1}^G{\bf \Omega}_g\right], \\
\bullet& \;\;\text{Compute the overall DAB as: } {\bf A}^* = \l[{\bf U}_{{\bf S}^{({\rm c})}}\r]_{1:L}^H {\bf A}_{({\rm i})}^*,
\end{aligned}
}
\end{equation}
where $\l[{\bf U}_{{\bf S}^{({\rm b})}}\r]_{1:R_{\rm s}}$ and $\l[{\bf U}_{{\bf S}^{({\rm c})}}\r]_{1:L}$ are the $R_{\rm s}$-dimensional and $L$-dimensional principal eigen-space of ${\bf S}^{({\rm b})}$ and ${\bf S}^{({\rm c})}$, respectively.

\begin{remark}[Cluster-based feedback] \emph{
In the case of overlapping clusters, even though the feedback is not one-shot and divided into separate feedback for different clusters, the feedback for devices within the same  cluster are simultaneous. Consequently, the total feedback overhead depends on the number of clusters and does not scale with the number of devices.}
\end{remark}

\section{Simulation Results}
Consider an IoT network with an AP and $G$ clusters of devices. There are $K$ devices in each cluster. The AP performs AirComp over the data transmitted by the devices. The simulation parameters are summarized in Table \ref{tb:Para}.
\begin{table}[!htp]
\caption{Simulation parameters}\label{tb:Para}
\centering{}%
\small
\begin{tabular}[t]{|l|l|}
\hline
Parameter & Value\tabularnewline
\hline
\hline
Bandwidth, {\it{W}} &${\rm{10~ MHz}}$\tabularnewline
\hline
Noise power density, $N_0$ &${\rm{-174~ dBm/Hz}}$\tabularnewline
\hline
Path loss between device and AP  &${\rm{145.4+37.5\log\left(0.05({\rm{\mathrm{km}}})\right)}}$\tabularnewline
\hline
Number of transmit antennas, $N_t$ & $5$\tabularnewline
\hline
Receive antenna spacing over wavelength, $D$ & $1/3$\tabularnewline
\hline
\end{tabular}
\end{table}

\subsection{Gains of Channel-Rank Selection}

\subsubsection{Channel-rank selection for disjoint clusters}
\begin{figure}[!htp]
    \centering
    \begin{subfigure}[b]{0.48\textwidth}
        \includegraphics[width=\textwidth]{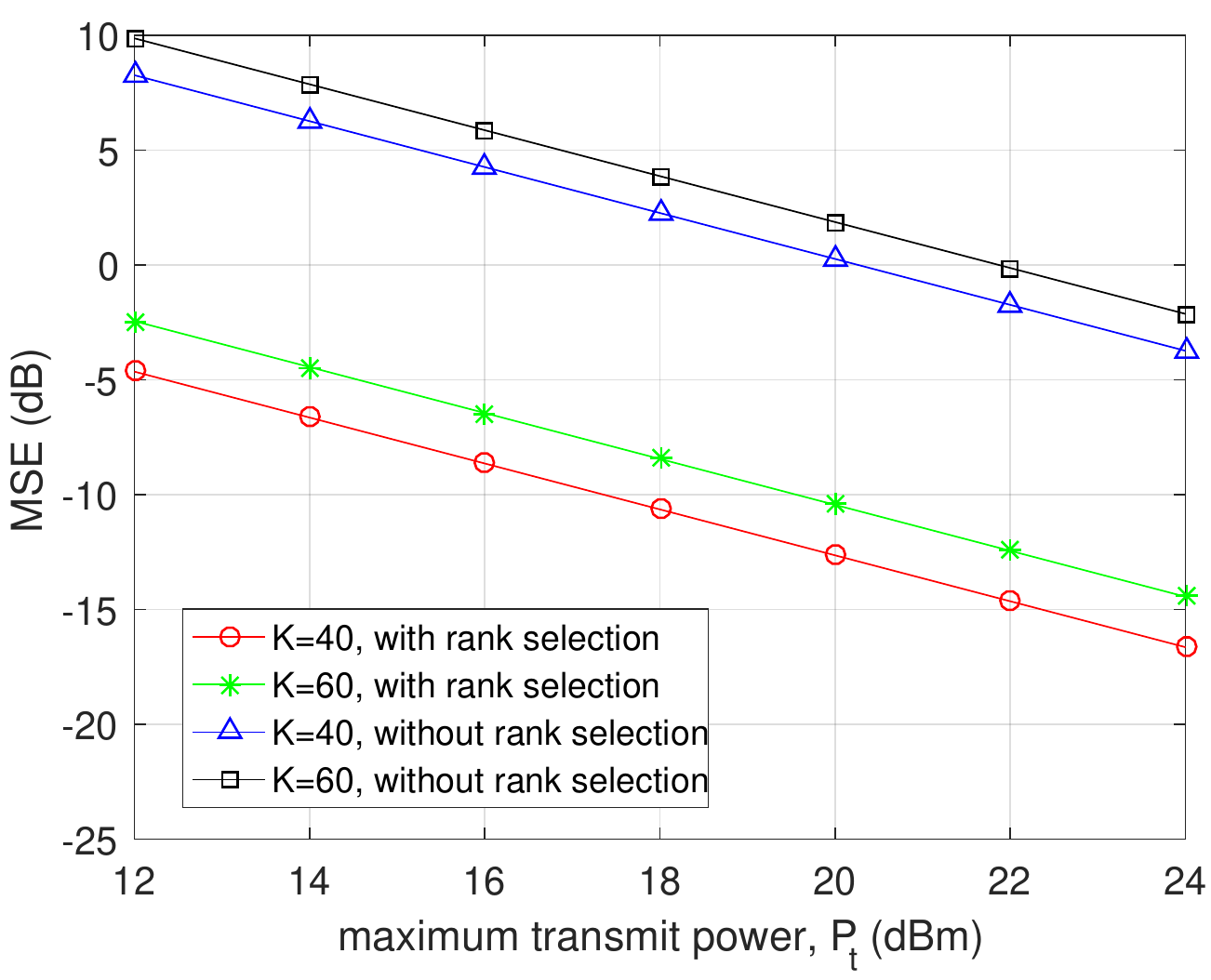}
        \caption{MSE with the maximum transmit power, $P_t$.}\label{fig:homP}
    \end{subfigure}
    ~
    \begin{subfigure}[b]{0.48\textwidth}
        \includegraphics[width=\textwidth]{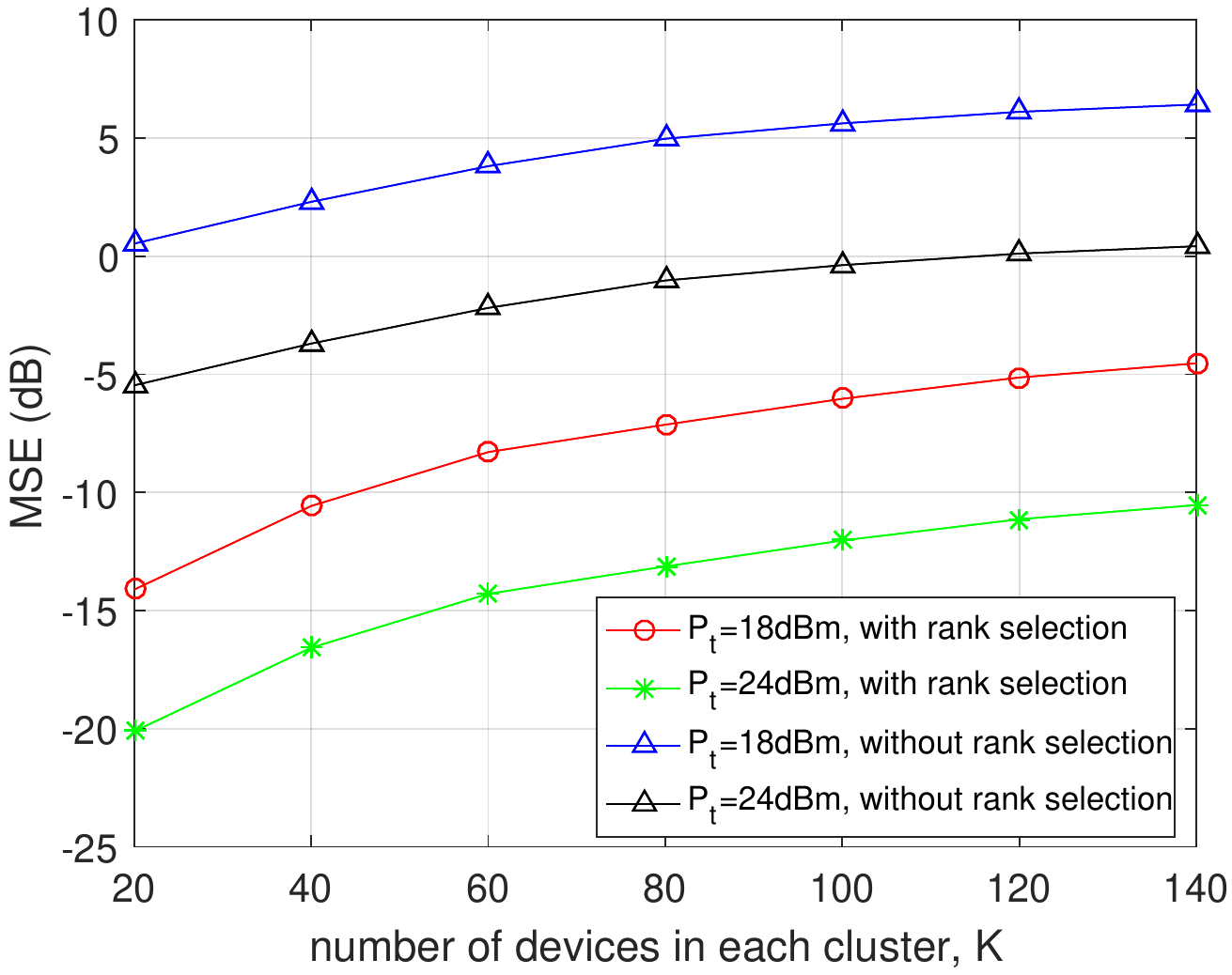}
        \caption{MSE with the number of devices in each cluster. }\label{fig:homU}
    \end{subfigure}
    \caption{Homogeneous channel-rank-selection for disjoint clusters.}\label{fig:hom}
\end{figure}

The homogeneous channel-rank selection of two disjoint clusters is presented in Fig. \ref{fig:hom}. In this case, the number of receive arrays is $N_r=48$. The AoA ranges are $\Delta \theta_1=[-49^{\circ},-1^{\circ}]$ and $\Delta \theta_2 = [1^{\circ},49^{\circ}]$, respectively. According to \cite{JSDM}, the number of ranks of both clusters can be calculated as $R_1=R_2=12$. In Fig. \ref{fig:hom}, the optimal homogeneous rank selection scheme can significantly improve the performance compared with no rank selection. Besides, Fig. \ref{fig:hom}(a) shows the MSE decreases with the maximum transmit power. The reason is that the MSE is inversely proportional to the maximum transmit power.
Fig. \ref{fig:hom}(b) shows the MSE increases with the number of devices in each cluster. The reason is that the sub-space distances between the effective channels after dimension reduction, on which the AirComp performance depends, increases with the number of devices.

\begin{figure}[!htp]
    \centering
    \begin{subfigure}[b]{0.48\textwidth}
        \includegraphics[width=\textwidth]{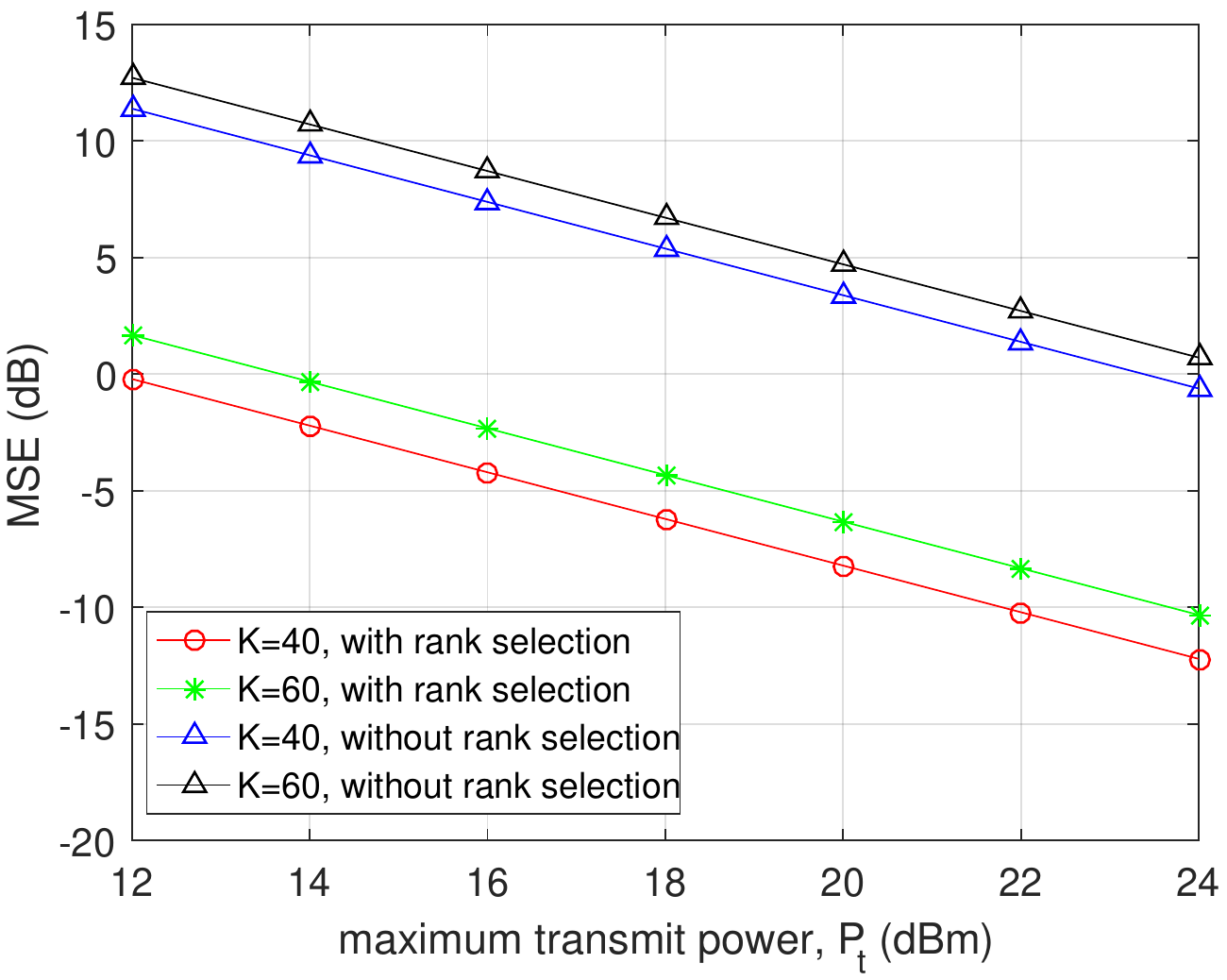}
        \caption{MSE with the maximum transmit power, $P_t$.}\label{fig:hetP}
    \end{subfigure}
    ~
    \begin{subfigure}[b]{0.48\textwidth}
        \includegraphics[width=\textwidth]{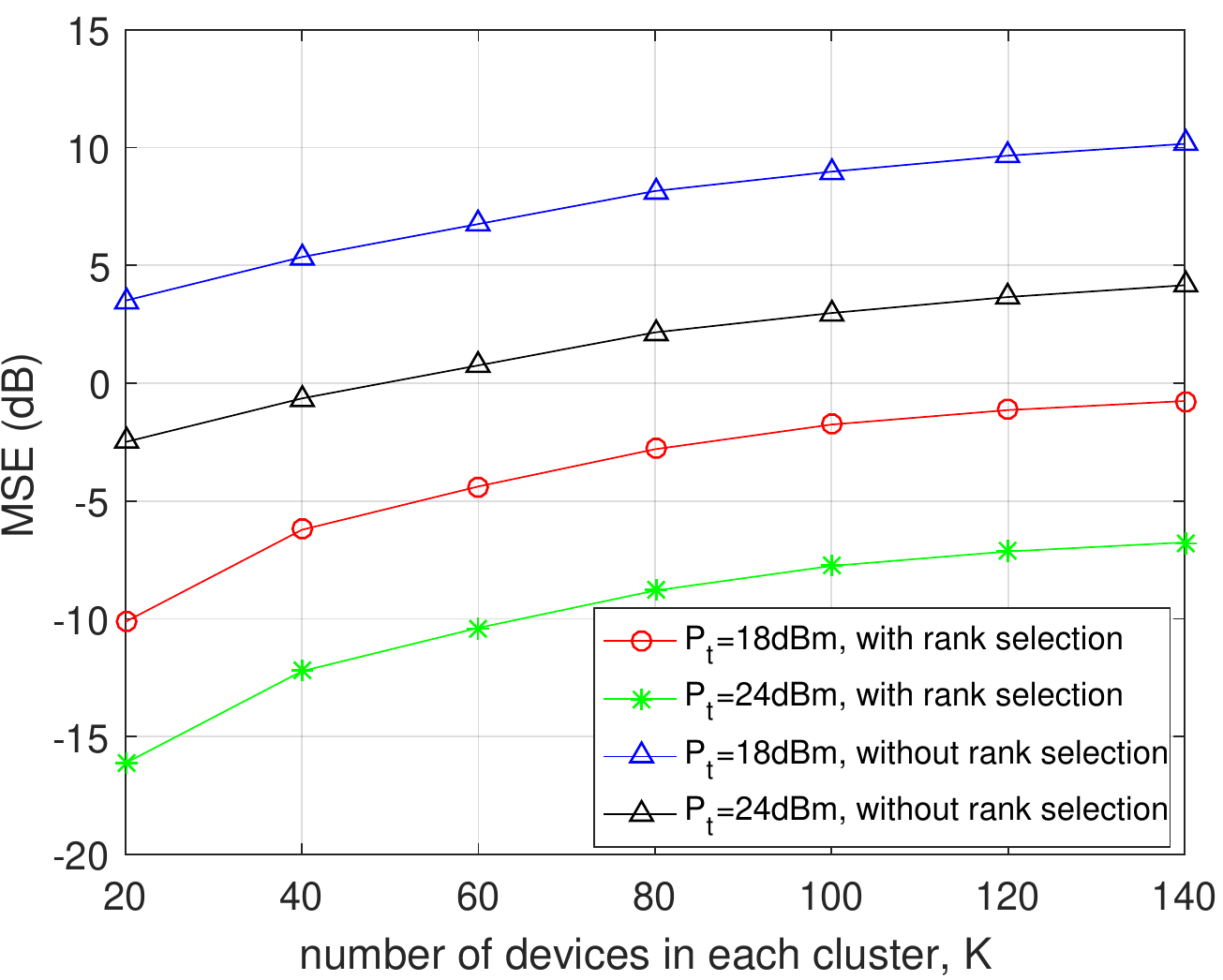}
        \caption{MSE with the number of devices in each cluster. }\label{fig:hetU}
    \end{subfigure}
    \caption{Heterogeneous channel-rank-selection for disjoint clusters. }\label{fig:het}
\end{figure}

The performance of the heterogenous channel-rank selection of three disjoint clusters is shown in Fig. \ref{fig:het}. In this case, the number of receive arrays is $N_r=48$. The AoA ranges are $\Delta \theta_1=[-51^{\circ},-15^{\circ}]$, $\Delta \theta_2 = [-14^{\circ},14^{\circ}]$, and $\Delta \theta_3 = [15^{\circ},41^{\circ}]$, respectively. The corresponding ranks are $R_1=8$, $R_2=8$, and $R_3=6$, respectively. In Fig. \ref{fig:het}, the sub-optimal
channel-rank-selection scheme proposed in Algorithm \ref{Ag:1} can improve the performance. Again, the MSE decreases with the maximum transmit power and increases with the number of devices in each cluster.

\subsubsection{Rank selection for overlapping clusters}
\begin{figure}[!htp]
    \centering
    \begin{subfigure}[b]{0.48\textwidth}
        \includegraphics[width=\textwidth]{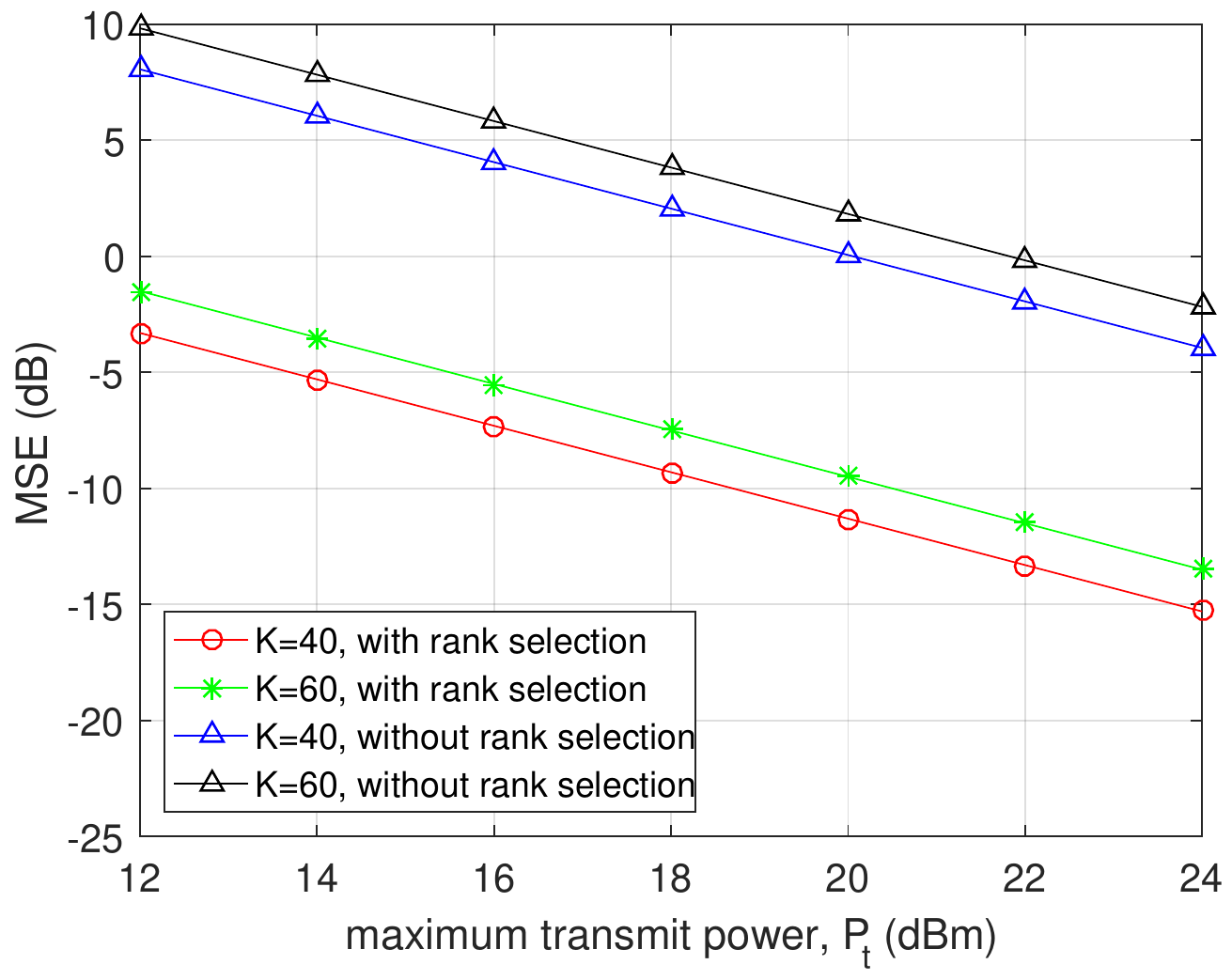}
        \caption{MSE with the maximum transmit power, $P_t$.}
    \end{subfigure}
    ~
    \begin{subfigure}[b]{0.48\textwidth}
        \includegraphics[width=\textwidth]{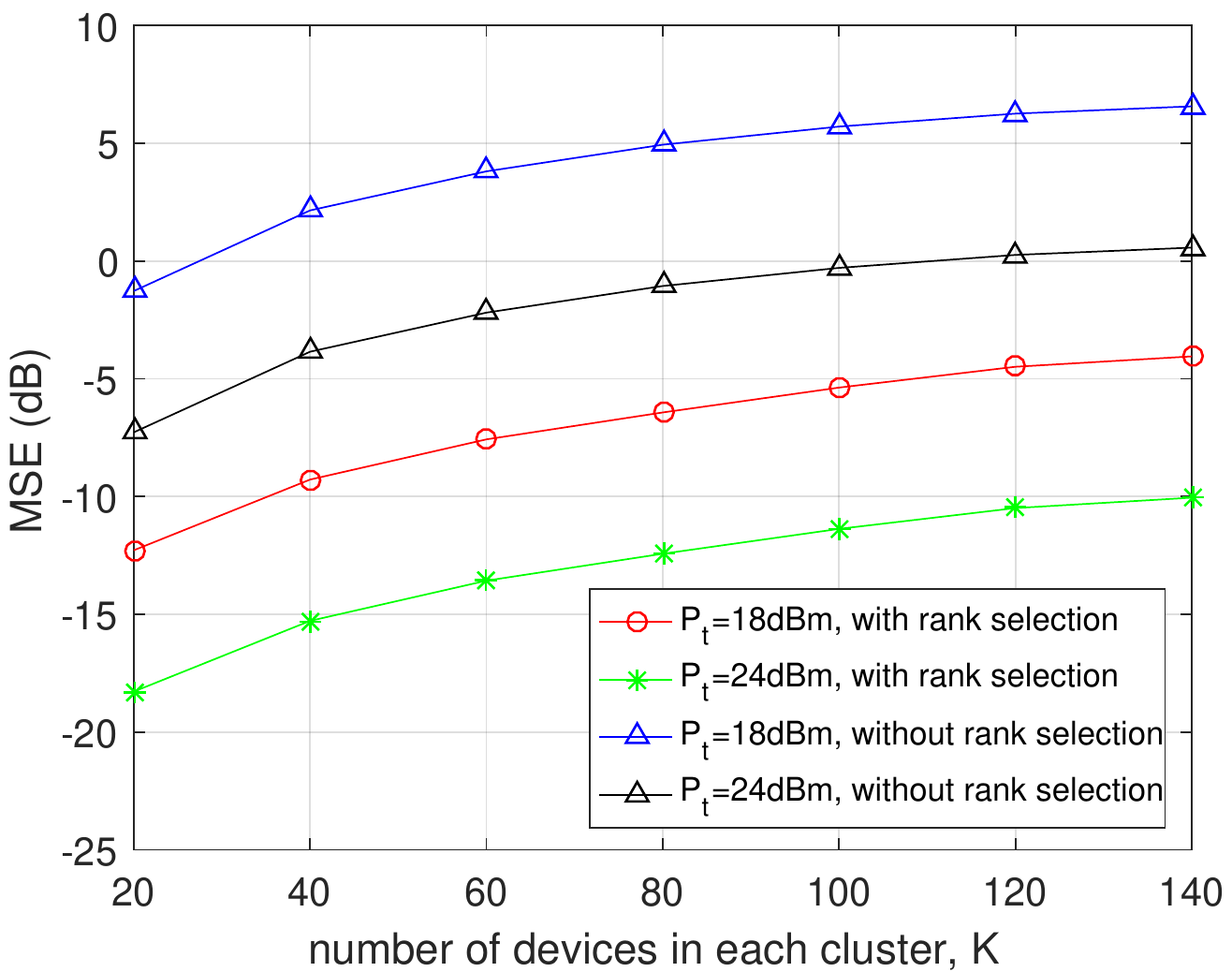}
        \caption{MSE with the number of devices in each cluster. }
    \end{subfigure}
    \caption{Channel-rank-selection for overlapping clusters. }\label{fig:Ovp}
\end{figure}

The performance of the channel-rank selection for overlapping clusters of two overlapping clusters is presented in Fig. \ref{fig:Ovp}. In this case, the number of receive arrays is $N_r=48$. The AoA ranges are $\Delta \theta_1=[-45^{\circ},15^{\circ}]$ and $\Delta \theta_2 = [-15^{\circ},45^{\circ}]$, respectively. The ranks of both clusters are $R_1=R_2=15$. Fig. \ref{fig:Ovp} shows that the channel-rank selection in this case can improve the performance.

The simulation results above verify that channel-rank selection is needed to achieve the best performance of AirComp in massive MIMO systems. Besides, increasing the transmit power and selecting the devices with most correlated small-scale fading can also improve the performance.

\subsection{Gains of Reduced-Dimension Aggregation Beamforming Design}
\begin{figure}[!htp]
    \centering
    \begin{subfigure}[b]{0.48\textwidth}
        \includegraphics[width=\textwidth]{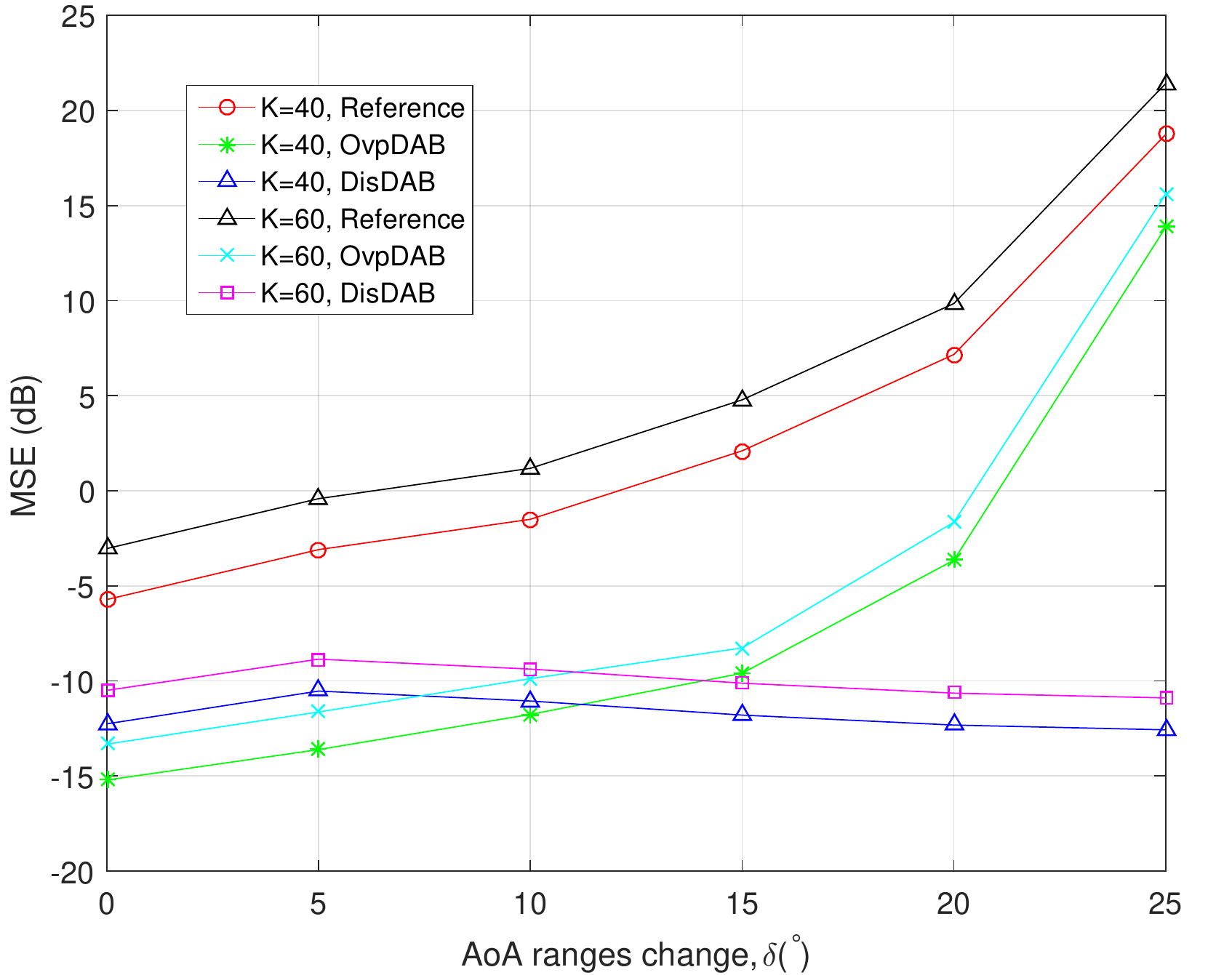}
        \caption{MSE with the AoA ranges change, $\delta$.}
    \end{subfigure}
    ~
    \begin{subfigure}[b]{0.48\textwidth}
        \includegraphics[width=\textwidth]{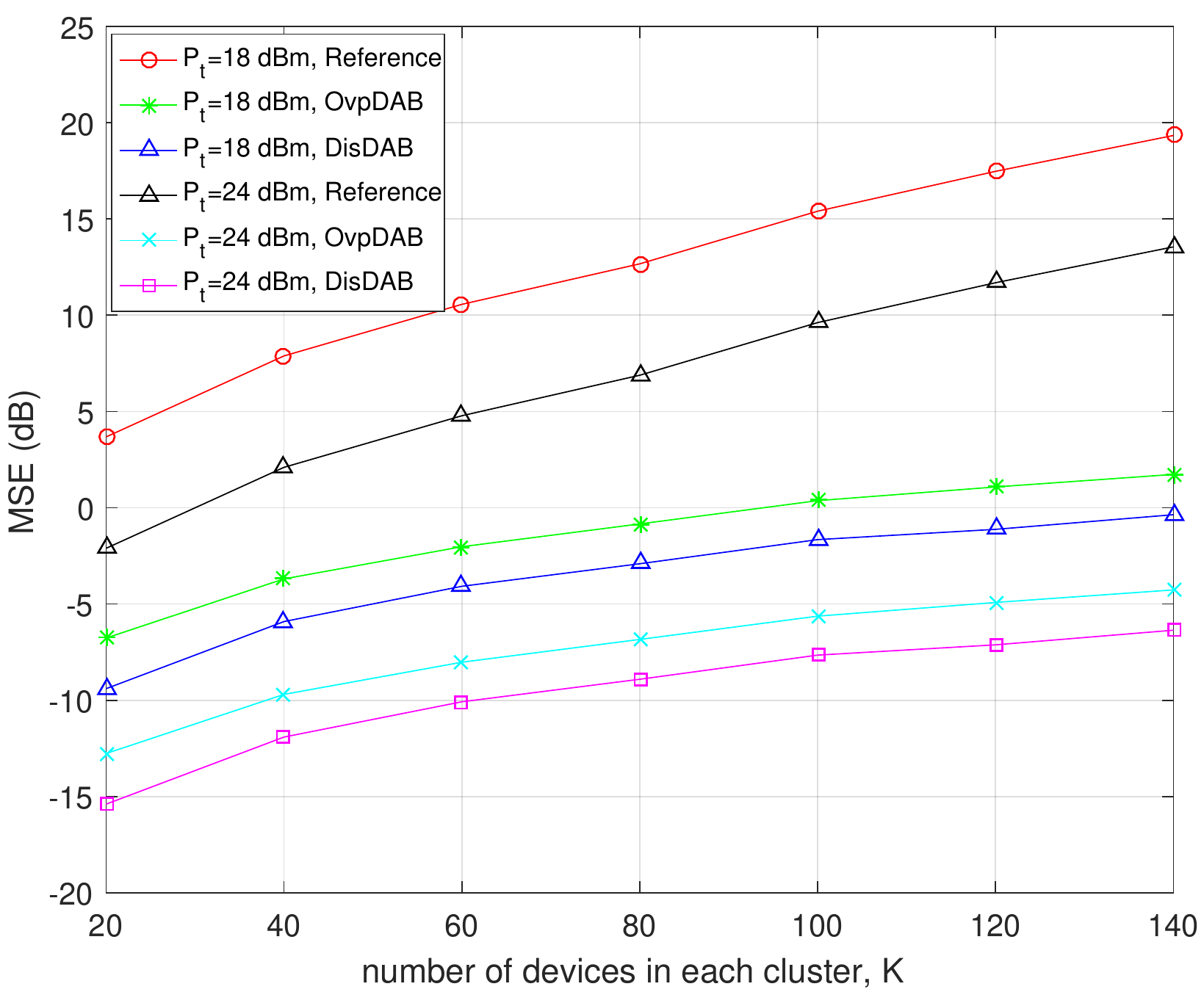}
        \caption{MSE with the number of users in each cluster. }

    \end{subfigure}
    \caption{Gains of reduced-dimension beamforming design.}\label{fig:cmp}
\end{figure}

In Fig. \ref{fig:cmp}, we show the gains of the proposed reduced-dimension design. Two clusters of devices are considered. To investigate the impact of channels' correlation, the number of receive arrays is set to $N_r=30$. In the figure, ``DisDAB", ``OvpDAB", and ``Reference" represents the beamforming design for disjoint clusters, overlapping clusters, and in \cite{Guangxu}, respectively.

In Fig. \ref{fig:cmp} (a), the transmit power is $P_t=24{\rm dBm}$. The AoA ranges of two clusters are $\Delta \theta_1= -\delta + [-35^{\circ},25^{\circ}]$, $\Delta \theta_2 = \delta + [-30^{\circ}, 30^{\circ}]$, respectively, where $\delta$ is the AoA ranges change. As $\delta$ increases, the AoA ranges of the two clusters changes from highly overlapping to nearly disjoint. In this figure, when the AoA ranges of the two clusters are highly overlapping, the DAB design for overlapping clusters has the best performance. Otherwise, the performance of the DAB design for disjoint clusters is the best. Besides, in highly overlapping case, the disjoint DAB design can still has good performance since the two clusters can nearly be regarded as one cluster.

In Fig. \ref{fig:cmp} (b), the AoA ranges are $\Delta \theta_1=[-50^{\circ},10^{\circ}]$, $\Delta \theta_2 = [-15^{\circ},45^{\circ}]$, respectively. It shows the the performance of DAB design for disjoint clusters is the best, because AoA ranges of the two clusters are not highly overlapping.

The simulation results above show that our proposed DAB designs can achieve better performance than the existing approach.

\section{Conclusion}
In this paper, we have presented the framework of reduced-dimension MIMO AirComp for clustered IoT networks. The design exploits the structure of clustered MIMO channel to reduce AirComp errors and channel-feedback overhead.  The key feature of the framework is the design of decomposed aggregation beamforming, which comprises  outer components performing channel dimension reduction and joint equalization of channel covariances and the inner components jointly equalize small-scale fading channels components.

The current work opens several directions for further investigation. One direction is  algorithmic design for MIMO AirComp. In particular,  sensor clustering algorithms can be designed to improve the performance of MIMO AirComp. Another interesting direction is to apply MIMO AirComp to  specific IoT or distributed-learning applications such as high-mobility UAV networks, federated learning or cloud coordinated vehicular platooning.

\begin{appendices}

\section{Proof Of Lemma \ref{lma:Decompose}}\label{apdx:lmaDCP}
We first construct an orthonormal basis of the $N_r$-dimensional space using the column vectors of $\l\{ {\bf U}_g, g\in[1,G] \r\}$. Then, each column vector of ${\bf A}$ is presented as a linear combination of the basis. Finally, the decomposition form of ${\bf A}$ is proved by combining the column vectors.

For notation simplicity, let $R_{G+1} = (N_r- \sum\nolimits_{i=1}^G R_g)$. Define a $N_r\times R_{G+1}$ dimensional unitary matrix, ${\bf U}_{G+1}$, which satisfies ${\bf U}_g^H{\bf U}_{G+1} = {\bf 0}$ for all $g\in[1,G]$. Then, the column vectors of $\l\{ {\bf U}_g, g\in[1,G+1] \r\}$ forms an orthonormal basis of the $N_r$-dimensional space. Denote the $i$-th column vector of ${\bf U}_g$ and ${\bf A}^H$ as ${\bf u}_{g,i}$ and ${\bf a}_i$, respectively. According to the projection theory, we have
\begin{equation}
{\bf a}_i = \sum\limits_{g=1}^{G+1}\sum\limits_{j=1}^{R_g}c^{(g)}_{i,j}{\bf u}_{g,j} = \sum\limits_{g=1}^{G+1} {\bf U}_g {\bf c}_{g,i}, \forall i\in[1,L],
\end{equation}
where $c_{i,j}^{(g)}$ the coefficient and ${\bf c}_{g,i} = \l[c_{i,1}^{g},c_{i,2}^{(g)},...,c_{i,R_g}^{(g)}\r]^T $, respectively. Thereby, ${\bf A}^H$ can be presented as
\begin{equation}
{\bf A}^H = \l[{\bf a}_1,{\bf a}_2,...{\bf a}_L\r] = \sum\limits_{g=1}^{G+1} {\bf U}_g \l[{\bf c}_{g,1}, {\bf c}_{g,2}, ..., {\bf c}_{g,L} \r] = \sum\limits_{g=1}^{G+1} {\bf U}_g {\bf C}_g,
\end{equation}
where ${\bf C}_g = \l[{\bf c}_{g,1}, {\bf c}_{g,2}, ..., {\bf c}_{g,L} \r]$.

Besides, with the channel, ${\bf H}_{g,k}$, defined in \eqref{eq:ChModel}, we have ${\bf C}_{G+1}^H{\bf U}_{G+1}^H{\bf H}_{g,k}={\bf 0},\forall g\in[1,G]$.  That's to say, the component, ${\bf C}_{G+1}^H{\bf U}_{G+1}^H$, of ${\bf A}$ has no contribution solve Problem ({\bf P2}). Then, let ${\bf C}_{G+1}^H={\bf 0}$ for simplicity. Hence, we have
\begin{equation}
{\bf A} = \sum\limits_{g=1}^G  {\bf C}_g^H {\bf U}_g^H.
\end{equation}
This completes the proof.

\section{Proof Of Lemma \ref{lma:EGA}}\label{apdx:lmaEGA}
Denoting ${\bf J}_{g,k}={\bf A}{\bf \hat{H}}_{g,k}{\bf \hat{H}}_{g,k}^H{\bf A}^H$ and substituting ${\bf A} = {\bf A}_{({\rm o})}{\bf A}_{({\rm i})}$ and \eqref{eq:ExModel}, we have
\begin{equation}
{\bf J}_{g,k} = {\bf A}_{({\rm o})}{\bf A}_{({\rm i})}{\bf \hat{U}}_g{\bf \hat{\Lambda}}_g^{\frac{1}{2}}{\bf \hat{W}}_{g, k}{\bf \hat{W}}_{g, k}^H{\bf \hat{\Lambda}}_g^{\frac{1}{2}}{\bf \hat{U}}_g^H{\bf A}_{({\rm i})}^H{\bf A}_{({\rm o})}^H,
\end{equation}
whose eigenvalues can be approximated to
\begin{equation}\label{eq:EGVJ2}
\begin{split}
\lambda_i\l( {\bf J}_{g,k} \r) &\geq \lambda_{\min}\l({\bf A}_{({\rm o})}{\bf A}_{({\rm i})}{\bf \hat{U}}_g {\bf \hat{U}}_g^H{\bf A}_{({\rm i})}^H{\bf A}_{({\rm o})}^H \r) \lambda_i\l( {\bf \hat{\Lambda}}_g^{\frac{1}{2}}{\bf \hat{W}}_{g, k}{\bf \hat{W}}_{g, k}^H{\bf \hat{\Lambda}}_g^{\frac{1}{2}} \r)  , \\
&\geq \lambda_{\min} \l({\bf A}_{({\rm i})}{\bf \hat{U}}_g {\bf \hat{U}}_g^H{\bf A}_{({\rm i})}^H \r) \lambda_{\min}\l({\bf A}_{({\rm o})}{\bf A}_{({\rm o})}^H \r) \lambda_i\l( {\bf \hat{\Lambda}}_g^{\frac{1}{2}}{\bf \hat{W}}_{g, k}{\bf \hat{W}}_{g, k}^H{\bf \hat{\Lambda}}_g^{\frac{1}{2}} \r),\\
& = \lambda_{\min} \l({\bf A}_{({\rm i})}{\bf \hat{U}}_g {\bf \hat{U}}_g^H{\bf A}_{({\rm i})}^H \r) \lambda_i\l( {\bf \hat{\Lambda}}_g^{\frac{1}{2}}{\bf \hat{W}}_{g, k}{\bf \hat{W}}_{g, k}^H{\bf \hat{\Lambda}}_g^{\frac{1}{2}} \r),
\end{split}
\end{equation}
where the last equality above is because ${\bf A}_{({\rm o})}{\bf A}_{({\rm o})}^H ={\bf I}$. Besides, $\tr \l(({\bf A}{\bf \hat{H}}_{g,k}{\bf \hat{H}}_{g,k}^H{\bf A}^H)^{-1}\r) = \sum\nolimits_{i=1}^L \lambda_i^{-1}\l( {\bf J}_{g,k} \r)$. By substituting \eqref{eq:EGVJ2}, we have
\begin{equation}
\tr \l(({\bf A}{\bf H}_{g,k}{\bf H}_{g,k}^H{\bf A}^H)^{-1}\r)\leq \lambda_{\min} \l({\bf A}_{({\rm i})}{\bf \hat{U}}_g {\bf \hat{U}}_g^H{\bf A}_{({\rm i})}^H \r)\sum\nolimits_{i=1}^{L} \lambda_i^{-1}\big( {\bf \hat{\Lambda}}_g^{\frac{1}{2}} {\bf \hat{W}}_{g,k}{\bf \hat{W}}_{g,k}^H {\bf \hat{\Lambda}}_g^{\frac{1}{2}} \big).
\end{equation}
This completes the proof.

\section{Proof of Lemma \ref{lma:EGVRank}}\label{apdx:lmaEGVRank}

By substituting ${\bf F}_{g,k}= {\bf \Lambda}_g^{\frac{1}{2}}{\bf W}_{g,k}$, we have $\lambda_{\min} \l({\bf F}_{g,k}^H{\bf F}_{g,k}\r) = \lambda_{\min}\l( {\bf W}_{g,k}^H{\bf \Lambda}_g{\bf W}_{g,k} \r)$. Besides, each element of ${\bf W}_{g, k}$ is i.i.d., and follows $\mathcal{CN}(0,1)$. For notation simplicity, let $r_g=r$. Let ${\bf T}_r = {\bf W}_{g,k,r}^H{\bf \Lambda}_{g,r}{\bf W}_{g,k,r}$ and ${\bf T}_{r+1} = {\bf W}_{g,k,r+1}^H{\bf \Lambda}_{g,r+1}{\bf W}_{g,k,r+1}$, where the size of ${\bf W}_{g,k,r}$ and ${\bf W}_{g,k,r+1}$ are $r\times N_t$ and $(r+1)\times N_t$, and ${\bf \Lambda}_{g,r}$ and ${\bf \Lambda}_{g,r+1}$ are the $r$ and $r+1$ dominant eigenvalue matrix of ${\bf \Lambda}_g$, respectively. Therefore, we have
\begin{equation}
{\bf T}_{r+1} = {\bf T}_r+ {\bf \Delta},
\end{equation}
where ${\bf \Delta}$ is a non-negative matrix. According to Weyl's inequality in matrix theory \cite{Weyl},
\begin{equation}
\lambda_i\l({\bf T}_r\r)\leq \lambda_i\l({\bf T}_{r+1}\r).
\end{equation}
Therefore, $\lambda_{\min}\l({\bf T}_r\r)\leq \lambda_{\min}\l({\bf T}_{r+1}\r)$. That's to say, $\lambda_{\min}\l({\bf F}_{gk}^H{\bf F}_{gk}\r)$ increases with $r$. This completes the proof.

\end{appendices}

{}

\end{document}